\documentclass[showpacs,preprintnumbers,amsmath,amssymb,twocolumn]{revtex4}

\usepackage{graphicx}
\usepackage{dcolumn}
\usepackage{bm}
\usepackage{graphicx}
\usepackage{amssymb}
\usepackage{amsmath}

\def\be{\begin{equation}}
\def\ee{\end{equation}}
\def\bea{\begin{eqnarray}}
\def\eea{\end{eqnarray}}

\newcommand{\G}{{Q}} 
\newcommand{\K}{{\rm K}} 
\newcommand{\set}[1]{{{[\rm #1]}}} 

\newcommand{\Veig}{{\nu}} 

\begin{document}

\title{O(6) Algebraic Theory Of Three Nonrelativistic Quarks Bound by Spin-independent Interactions}
\author{V. Dmitra\v sinovi\' c}
\author{Igor Salom}
\affiliation{Institute of Physics, University of Belgrade, Pregrevica 118, Zemun, \\
P.O.Box 57, 11080 Beograd, Serbia }

\date{\today}
%
\begin{abstract}
We apply the newly developed theory of permutation-symmetric O(6) hyperspherical harmonics
to the quantum-mechanical problem of three non-relativistic quarks confined by a spin-independent
3-quark potential. We use our previously derived results to reduce the three-body Schr\"odinger equation
to a set of coupled ordinary differential equations in the hyper-radius $R$ with coupling coefficients
expressed entirely in terms of (i) a few interaction-dependent O(6) expansion coefficients and (ii) O(6)
hyperspherical harmonics matrix elements, that have been evaluated in our previous paper.
This system of equations allows a solution to the eigenvalue problem with homogeneous 3-quark potentials,
which class includes a number of
standard Ans\"atze for the confining potentials, such as the Y- and $\Delta$-string ones.
We present analytic formulae for the $\K = 2, 3, 4, 5$ shell states' eigen-energies in homogeneous
three-body potentials, which formulae we then apply to the Y- and $\Delta$-string, as well as the logarithmic
confining potentials. We also present numerical results for power-law pair-wise potentials with the
exponent ranging between -1 and +2. In the process we resolve the 25 year-old Taxil \& Richard vs. Bowler et al.
controversy regarding the ordering of states in the $\K = 3$ shell, in favor of the former.
Finally, we show the first clear difference between the spectra of
$\Delta$- and Y-string potentials, which appears in $\K \geq 3$ shells.
Our results are generally valid, not just for confining potentials, but also for many momentum-independent
permutation-symmetric homogenous potentials, that need not be pairwise sums of two-body terms.
The potentials that can be treated in this way must be square-integrable under
the O(6) hyperangular integral, however, which class does not include the Dirac $\delta$-function.
\end{abstract}

\maketitle

\section{Introduction}
\label{s:Introduction}

The nonrelativistic three-quark system has been the basis of our understanding of baryon
spectroscopy for more than 50 years; of course, this model has also many limitations, its
nonrelativistic character being just one of several. After the November 1974 
discovery of charmed hadrons, the nonrelativistic nature stopped being a detriment, 
at least in the case of heavy quarks. There are, of course, still only comparatively few
heavy-quark baryons in Particle Data Group (PDG) tables, and fewest of all, the triple-heavy ones.
That circumstance will not prevent us from trying to understand them, however. Indeed, even if there
were no heavy-quark baryons at all, it would still be an important systematic question to answer, if
for no other reason than to have a definite benchmark against which to compare relativistic calculations.

Chronologically, at first all calculations were done with a harmonic oscillator potential, due to its
integrability,
but with passing time other, ``more realistic'' potentials, such as the pair-wise sum of the Coulomb
and linearly rising two-body potentials plus various forms of ``strong hyperfine'' interactions,
have been used in numerical calculations. Such calculations generally involve uncontrolled,
sometimes drastic approximations, such as the introduction of cutoff(s),
due to the contact nature of the ``strong hyperfine'' interactions, thus leaving open many questions
about the level-ordering, convergence, and even existence of energy spectra in such calculations
\footnote{there are no mathematical theorems guaranteeing existence of a well defined spectrum in
the three-body problem \cite{Grosse:1997xu}, as there are in
the two-body problem, Refs. \cite{Grosse:1979xm,Baumgartner:1984cx,Martin:1989mm,Grosse:1997xu}.
Several attempts at a mathematically well-defined theory of nonrelativistic three-quark systems
have been recorded \cite{Bowler:1981xh,Martin:1989mm,Richard:1989ra}, but to little avail.}.

In this, the third in a series of papers, we show that the nonrelativistic three-quark
problem does have a well-defined spectrum for a class of (homogeneous) potentials that includes the
``standard'' confinement potentials. 
This development is based on two previous (sets of) papers: (1) Refs. \cite{Salom:2016vcw,Salom:2017ekb}
wherein the three-body permutation symmetry adapted O(6) hyperspherical were constructed; and
(2) Ref. \cite{Salom:2016pla} where we applied the said permutation symmetry adapted O(6) hyperspherical
harmonics to the problem of three nonrelativistic identical particles in a homogeneous potential.
Here we present a mathematically well-defined 
method for solving the three-heavy-quark problem, together with several examples: the $\K = 0, \cdots,5$ shells.
These examples turn out to be (very) instructive, as they clearly mark out the region of applicability of our
method.

In spite of the huge literature on the quantum-mechanical three-body bound-state problem,
in which the hyperspherical harmonics play a prominent role,
Refs. \cite{DELVES:1958zz,Smith:1959zz,Simonov:1965ei,Iwai1987a},
there are still many open problems related to the general structure of the three-body bound-state
spectrum (e.g.\ the ordering of states, even in the simplest case of three identical
particles). \footnote{In comparison, the two-body bound state problem is well understood,
see Refs. \cite{Grosse:1979xm,Baumgartner:1984cx,Martin:1989mm,Grosse:1997xu},
where theorems controlling the ordering of bound states in convex two-body potentials
were proven more than 30 years ago.} The core of the existing difficulties can be traced
back to the absence of a systematic construction of permutation-symmetric three-body wave
functions. Until recently, see Refs. \cite{Salom:2016vcw,Salom:2017ekb},
permutation symmetric three-body hyperspherical harmonics in three dimensions were
known explicitly only in a few special cases, such as those with total orbital angular
momentum $L=0,1$, Refs. \cite{Simonov:1965ei,Barnea:1990zz}.

In this paper we confine ourselves to the study of factorizable (into hyper-radial and hyper-angular parts)
three-body potentials that are square-integrable\footnote{This is not the first time square-integrability
of the potential has been demanded in quantum mechanics: Kato used it in his study of
``Eigenfunctions of Many-Particle Systems in Quantum Mechanics'', Ref. \cite{Kato:1957}.}
(in hyper-angles) for technical reasons: For this class of potentials our method allows closed-form
(``analytical'') results, at sufficiently small values of the grand angular momentum $\K$ (i.e. up to,
and including the $\K \leq 8$ shell). Factorizable potentials include homogenous potentials,
which in turn include pair-wise sums of two-body power-law potentials, such as the linear (confining)
``$\Delta$-string'', the 
``Y-string'' potential \cite{Dmitrasinovic:2009ma,Dmitrasinovic:2014}, as well as the Coulomb one.
Lattice QCD studies \cite{Takahashi:2002bw,Alexandrou:2002sn,Koma:2017hcm} suggest that
three-static-quarks potential is a (linear) combination of the aforementioned three.

Singular potentials, such as the (strong, or electromagnetic) hyperfine interactions, that include
the Dirac $\delta$-function, even though homogeneous, do not fall into the class of potentials
susceptible to this method, as they are not square-integrable - therefore they require special attention,
and will be treated elsewhere.
The spin-orbit potentials generally involve both the spin and the spatial variables for their
permutation invariance, which requires special techniques.
Simple inhomogenous potentials can only be treated numerically, however, using our method.

Strictly speaking, our (present) results are applicable only to three-equal-heavy-quark systems,
of which not one has been created in experiment, thus far (which does not mean that some
are not forthcoming). This condition limits the method's applicability to $c^3$ and $b^3$ baryons only.
Of course, in these two cases there is no flavour multiplicity, and we may drop
the $SU_{FS}(6)$ and $SU_{F}(3)$ labels. Nevertheless, we have kept the full $SU_{FS}(6)$ and $SU_{F}(3)$
labels, in the hope that in the future the present methods can and will be extended to
(a) two-identical and one distinct heavy quark systems, such as the $c^2 b$ and $b^2 c$; and
(b) (semi)relativistic three-light-quark systems.


This paper is divided into six Sections and one Appendix. After the present Introduction,
in Sect. \ref{s:HS3D} we show how the Schr\" odinger equation for three particles in a
homogenous/factorizable potential
can be reduced to a single differential equation and an algebraic/numerical problem
for their coupling strengths.
In Sect. \ref{s:potential} we defined the Y- and $\Delta$-string, the QCD Coulomb,
and the Logarithmic potential, and calculated the four lowest O(6)
hyperspherical harmonics expansion coefficients, that are relevant to $\K \leq 5$ shell states.
In Sect. \ref{s:Results} calculate the $\K=2,3,4,5$ shells' 
level splittings in terms of four parameters that characterize the three-body potential.
In Sect. \ref{s:Discussion} we discuss our results,
and 
in Sect. \ref{s:Summary} we summarize and draw conclusions.
The details of calculations are shown in Appendix \ref{a:calculations}.

\section{Three-body problem in hyper-spherical coordinates}
\label{s:HS3D}

In this section we shall closely follow the treatment of the nonrelativistic three-body
problem presented in Ref. \cite{Salom:2016pla}.

The three-body wave function $\Psi({\bm \rho},{\bm \lambda})$
can be transcribed from the Euclidean relative position (Jacobi) vectors
${\bm \rho} = \frac{1}{\sqrt{2}}({\bf x_1} - {\bf x_2})$,
${\bm \lambda} = \frac{1}{\sqrt{6}}({\bf x_1} + {\bf x_2}- 2 {\bf
x_3}),$
into hyper-spherical coordinates as $\Psi(R, \Omega_5)$, where
$R = \sqrt{{\bm \rho}^2 + {\bm \lambda}^2}$ is the hyper-radius,
and five angles $\Omega_5$ that parametrize a hyper-sphere in the
six-dimensional Euclidean space. Three ($\Phi_i;~ i=1,2,3$) of these five
angles ($\Omega_{5}$) are just
the Euler angles associated with the orientation in a three-dimensional
space of a spatial reference frame defined by the (plane of) three bodies; the remaining
two hyper-angles describe the shape of the triangle subtended
by three bodies; they are functions of three independent scalar three-body variables, e.g.
${\bm \rho} \cdot {\bm \lambda}$, ${\bm \rho}^2$, and ${\bm \lambda}^2$.
As we saw above, one linear combination of the two variables ${\bm \rho}^2$,
and ${\bm \lambda}^2$, is already taken by the hyper-radius $R$, so the
shape-space is two-dimensional, and topologically equivalent to
the surface of a three-dimensional sphere.

There are two traditional ways of parameterizing this sphere:
1) the standard Delves choice, \cite{DELVES:1958zz}, of hyper-angles $(\chi, \theta)$,
that somewhat obscures the full $S_3$ permutation symmetry of the problem;
2) the Iwai, Ref. \cite{Iwai1987a}, hyper-angles $(\alpha$, $\phi)$:
$(\sin \alpha)^2 = \,1 - \,\left(\frac{2 {\bm \rho} \times {\bm \lambda}}{R^{2}}\right)^{2}$,
$\tan \phi = \left(\frac{2 {\bm \rho} \cdot {\bm \lambda}}{{\bm \rho}^2
- {\bm \lambda}^2}\right)$, 
reveal the full $S_3$ permutation symmetry of the problem:
the angle $\alpha$ does not change under permutations, so that all permutation
properties are encoded in the $\phi$-dependence of the wave functions.
We shall use the latter choice, as it leads to permutation-symmetric hyperspherical
harmonics, as explained in Ref. \cite{Salom:2016vcw,Salom:2017ekb}
where specific hyperspherical harmonics used here are displayed.

We expand the wave function $\Psi(R, \Omega_5)$ in terms of hyper-spherical
harmonics ${\cal Y}^\K_{\set m}(\Omega_5)$,
$\Psi(R, \Omega_5) =  \sum_{\K, \set m} \psi^\K_\set m(R)
{\cal Y}^\K_\set m(\Omega_5)$,
where $\K$ together with $\set m = [\G, \Veig, L, L_z = m]$ constitute the complete set of
hyperspherical quantum numbers: $\K$ is the hyper-spherical angular 
momentum, $L$ is the (total orbital) angular momentum,
$L_z = m$ its projection on the z-axis, $\G$ is the Abelian quantum number conjugated with
the Iwai angle $\phi$, and $\Veig$ is the multiplicity label that
distinguishes between hyperspherical harmonics with remaining four quantum numbers
that are identical, see  Ref. \cite{Salom:2016vcw,Salom:2017ekb}.

The hyper-spherical harmonics turn the Schr\" odinger equation into a set of (infinitely)
many coupled equations,
\begin{eqnarray}
&-& \frac{1}{2\mu}\left[\frac{d^2}{dR^2} + \frac{5}{R}\frac{d}{dR} -
\frac{\K(\K+4)}{R^2} + 2\mu E \right]\psi^\K_\set m(R)\nonumber \\
&+&  
V_{\rm eff.}(R) \sum_{\K', \set{m'}} C^{\K \ \, \K'}_{\set m \set{m'}}\psi^{\K'}_\set{m'}(R) = 0
\label{e:Schrodinger}
\end{eqnarray}
with a hyper-angular coupling coefficients matrix $C^{\K \ \, \K'}_{\set m \set{m'}}$ defined by
\begin{eqnarray}
V_{\rm eff.}(R) C^{\K' \ \, \K}_{\set {m'} \set{m}} &=&
\langle {\cal Y}^{\K'}_\set{m'}(\Omega_{5})| V(R,\alpha,\phi) |
{\cal Y}^{\K}_\set{m}(\Omega_{5}) \rangle \nonumber \\
&=&  V_{}(R) \langle
{\cal Y}^{\K'}_\set{m'}(\Omega_{5})| V(\alpha, \phi)
|{\cal Y}^{\K}_\set{m}(\Omega_{5}) \rangle
\label{e:hyperang_ME2} . \
\end{eqnarray}
Factorizability of the potential is a simplifying assumption, that leads to analytic results
in the energy spectrum. It holds for several physically interesting potentials, such as
power-law ones, but also other homogeneous ones, see Sect. \ref{s:potential}. Unfortunately,
the sum (and difference) of two factorizable potentials is generally not factorizable itself.

In Eq.\ (\ref{e:Schrodinger}) we used the factorizability of the potential
$V(R,\alpha,\phi) = V(R) V(\alpha,\phi) $ to reduce this set to one (common) hyper-radial
Schr\" odinger equation.
The hyper-angular part $V\left(\alpha,\phi \right)$ can be expanded in terms of O(6) hyper-spherical
harmonics with zero angular momenta $L=m=0$ (due to the rotational invariance of the potential),
\begin{equation}
V\left(\alpha,\phi \right) =
\sum_{\K,\G}^{\infty} v_{\K,\G}^{\rm 3-body} {\cal Y}^{\K \G \Veig}_{00}(\alpha, \phi)
\label{e:Vdecomposition}
\end{equation}
where
\begin{equation}
v_{\K,\G}^{\rm 3-body} =
\int~ {\cal Y}^{\K\G\Veig *}_{00}(\Omega_{5}) \,  V
(\alpha, \phi)\, \, d\Omega_{(5)}
\label{e:def_ME}
\end{equation}
leading to
\begin{eqnarray}
V_{\rm eff.}(R) C^{\K'' \ \, \K'}_{\set {m''} \set{m'}}
=&&  V_{}(R) \sum_{\K,\G}^{\infty} v_{\K,\G}^{\rm 3-body} \nonumber \\
\langle {\cal Y}^{\K''}_\set{m''}(\Omega_{5})|&& {\cal Y}^{\K\G\Veig}_{00}(\alpha, \phi)
|{\cal Y}^{\K'}_\set{m'}(\Omega_{5}) \rangle
\label{e:hyperang_ME3} \
\end{eqnarray}
There is no summation over the multiplicity index in Eq.\ (\ref{e:Vdecomposition}),
because no multiplicity arises for harmonics with $L<2$.
Here we separate out the $\K=0$ term and absorb the factor
$\frac{v_{00}^{\rm 3-body}}{\pi \sqrt{\pi}} $ into the definition of
$V_{\rm eff.}(R) = \frac{v_{00}^{\rm 3-body}}{\pi \sqrt{\pi}} V_{}(R)$
to find
\begin{eqnarray}
& C^{\K'' \ \, \K'}_{\set {m''} \set{m'}} =
\delta_{\K'',\K'}\delta_{\set{m''},\set{m'}}  +
\pi\sqrt{\pi} \sum_{\K > 0,\G}^{\infty}
\frac{v_{\K,\G}^{\rm 3-body}}{v_{00}^{\rm 3-body}} &\nonumber \\
&\times
\langle {\cal Y}^{\K''}_\set{m''}(\Omega_{5})  
| {\cal Y}^{\K\G\Veig}_{00}(\alpha, \phi)
|{\cal Y}^{\K'}_\set{m'} (\Omega_{5}) 
\rangle. &
\label{e:hyperang_ME5} \
\end{eqnarray}

Homogenous potentials, such as the $\Delta$ and Y-string ones, which are linear in $R$,
and the Coulomb one, 
see Sect. \ref{s:potential} for definition of these potentials,
have the first coefficient $v_{00}^{\rm 3-body}$ in the h.s. expansion that is generally (at least)
one order of magnitude larger than the rest $v_{\K>0,\G}^{\rm 3-body}$, see
Table \ref{tab:Potential_expansion}, and
Fig. \ref{f:PowerPotDecomposition}.
This reflects the fact that, on the average, these potentials depend more on the overall size of the
system than on its shape, thus justifying the adiabatic (perturbative) approach taken in
Ref. \cite{Richard:1989ra}, with the first term in Eq.\ (\ref{e:hyperang_ME5}) taken as the
zeroth-order approximation.\footnote{Note that the h.s. matrix elements
$\langle {\cal Y}^{\K''}_\set{m''}(\Omega_{5})| {\cal Y}^{\K\G\Veig}_{00}(\alpha, \phi)
|{\cal Y}^{\K'}_\set{m'} (\Omega_{5}) \rangle$ under the sum are always less than $\frac{1}{\pi\sqrt{\pi}}$.}

In such cases Eqs. (\ref{e:Schrodinger}) decouple, leading to zeroth order solutions for
$\psi^{\ \K}_{0\set m}(R) $ that are independent of $[m]$ and thus have equal energies
within the same $\K$ shell, and different energies in different $\K$ shells.
Two known exceptions are potentials with the homogeneity degree $\rm k = -1, 2$,
that lead to ``accidental degeneracies'' and have to be treated separately.

The first-order corrections are obtained by diagonalization of the block matrices
$C^{\K \ \, \K}_{\set {m} \set{m'}} $, $\K=1,2,...$, while the off-diagonal couplings
$C^{\K \ \, \K'}_{\set {m} \set{m'}}, \K \neq \K'$ appear only in the second-order corrections.
Rather than calculating perturbative first-order energy shifts, a better approximation is 
obtained when the diagonalized block matrices are plugged back into Eq. \ (\ref{e:Schrodinger}),
which equations then decouple into a set of (separate) individual ODEs in one variable, that differ
only in the value of the effective coupling constant:
\begin{equation}
\left[\frac{d^2}{dR^2} + \frac{5}{R}\frac{d}{dR} -
\frac{\K(\K+4)}{R^2} + 2\mu (E - V^\K_{\set{m_d}}(R)) \right]\psi^\K_{\set{m_d}}(R) = 0,
\label{e:SchrodingerDecoupled}
\end{equation}
where $V^\K_{\set{m_d}}(R) = C^\K_{\set{m_d}}  V_{\rm eff.}(R)$, with $C^\K_{\set{m_d}}$ being the
eigenvalues of matrix $C^{\K \ \, \K}_{\set {m} \set{m'}} $.

The spectrum of three-body systems in homogenous potentials,
such as those considered in Refs. \cite{Salom:2016vcw,Salom:2017ekb},
is now reduced to finding the eigenvalues of a single differential operator, 
just as in the two-body problem with a radial potential.
The matrix elements in Eq. (\ref{e:hyperang_ME5}) can be readily evaluated using the permutation-symmetric
O(6) hyper-spherical harmonics and the integrals that are spelled out in  Refs. \cite{Salom:2016vcw,Salom:2017ekb}.

This is the main (algebraic) result 
of this section: combined with
the hyperspherical harmonics recently
obtained in Ref. \cite{Salom:2016vcw,Salom:2017ekb}, it allows one to evaluate
the discrete part of the (energy) spectrum of a three-body potential as a
function of its shape-sphere harmonic expansion coefficients $v_{\K,\G}^{\rm 3-body}$.
Generally, these matrix elements obey selection rules:
they are subject to the ``triangular'' conditions
$\K' + \K'' \geq \K \geq |\K' - \K''|$ plus the condition that
$\K' + \K'' + \K = 0, 2, 4, \dots$, and 
the angular momenta satisfy the 
selection rules: $L' = L''$, $m' = m''$.
Moreover, $\G$ is an Abelian (i.e. additive)
quantum number that satisfies the simple selection rule:
$\G'' = \G' + \G$.
That reduces the sum in Eq. (\ref{e:hyperang_ME5}) to a finite one, 
that depends on a finite number of coefficients $v_{\K,\G}^{\rm 3-body}$; for
small values of $\K$, this number is also small.

A matrix such as that in Eq. (\ref{e:hyperang_ME5}) is generally sparse in the
permutation-symmetric basis, so its diagonalization is not a serious 
problem, and, for sufficiently small $\K$ values it can even be accomplished
in closed form: for example, for $\K \leq 5$, all
results depend only on four coefficients $(v_{00}, v_{40}, v_{6\pm 6}, v_{80})$,
and there is at most three-state mixing, so the eigenvalue equations are at most cubic
ones, with well-known solutions.
As there is only a small probability that many states from the $\K \geq 6$ shells will
be observed in the foreseeable future,
we limit ourselves to $\K \leq 5$ shells here.

\section{Three-body spin-independent potentials}
\label{s:potential}

\subsection{The Lattice QCD three-static-quarks potential}
\label{s:LQCD_potential}

Lattice QCD calculations indicate that the confining interactions among quarks do not
depend on the quarks' spin and flavour degrees of freedom.

There have been several attempts at extracting the three-quark potential from lattice QCD
over the years, see Refs. \cite{Takahashi:2002bw,Alexandrou:2002sn,Koma:2017hcm}. They
were based on lattices of different sizes, $12^3 \times 24$ at $\beta =5.7$ and $16^3 \times 32$
at $\beta = 5.8, 6.0$ in Ref. \cite{Takahashi:2002bw}, $16^3 \times 32$ at $\beta = 5.8, 6.0$
in Ref. \cite{Alexandrou:2002sn}, and $24^4$ at $\beta =5.7, 5.8, 6.0$ in Ref. \cite{Koma:2017hcm}.
Moreover, Refs. \cite{Takahashi:2002bw,Alexandrou:2002sn} use the Wilson loop techniques,
whereas Ref. \cite{Koma:2017hcm} uses the Polyakov loop.
Their conclusions also differ markedly: Ref. \cite{Takahashi:2002bw} ``supports the Y {\it Ansatz}'',
Ref. \cite{Alexandrou:2002sn} ``finds support for the $\Delta$ {\it Ansatz}'', whereas the most recent
Ref. \cite{Koma:2017hcm} finds that
``The potentials of triangle geometries are clearly different from the half of the sum of the two-body
quark-antiquark potential'', i.e., suggesting that is not the $\Delta$ {\it Ansatz}.
All of which indicates that the lattice QCD potential is neither pure Y {\it Ansatz}, nor a pure
$\Delta$ {\it Ansatz}.

A detailed analysis \cite{Leech:2017} of the Ref. \cite{Takahashi:2002bw} and Ref. \cite{Koma:2017hcm} 
published data
in terms of hyperspherical coordinates has shown that these two groups have calculated the potential (mostly)
in very different geometric configurations, whose overlap is small so that neither calculation is conclusive.

It stands to reason that the definitive QCD prediction is a linear superposition of the two {\it Ans\"atze}
and the QCD Coulomb term,
but at this stage it is impossible to evaluate the lattice QCD potential's O(6) expansion coefficients
due to the dearth of evaluated points on the hypersphere.

For this reason, we shall analyze both {\it Ans\"atze}, separately, in addition to the QCD Coulomb potential,
which is a must. Finally, we shall also consider the logarithmic potential, which can be thought of
as the best homogeneous-potential approximation to the sum of the Coulomb and the linearly rising potential.

As stated in Sect. \ref{s:HS3D} above, any spin-independent three-body potential must be invariant
under overall (ordinary) rotations, as it is a scalar, i.e., it contains only the
zero-angular momentum hyperspherical components, which significantly simplifies the expansion of
the potential in O(6) hyperspherical harmonics.
Below we shall calculate these expansion coefficients in several homogeneous potentials. 

\subsection{The Y-string and other area-dependent potentials}
\label{s:Y_potential}

The complexity of the full Y-string potential, defined by
\begin{equation}
\label{conf_Y} V_{\rm Y-string} = \sigma_Y \min_{\bf x_0}\; \sum_{i=1}^3 |{\bf
x_i} - {\bf x_0}|.
\end{equation}
can best be seen when expressed in terms of three-body Jacobi (relative)
coordinates ${\bm \rho},{\bm \lambda}$, as follows.
The full Y-string potential  Eq. (\ref{conf_Y}),
consists of the so-called central Y-string, or ``Mercedes-Benz-string'' term,
\begin{eqnarray}
V_{\rm Y-central} &=&  \sigma_{\rm Y} \sqrt{\frac{3}{2}({\bm
\rho}^2 + {\bm \lambda}^2 + 2 |{\bm \rho} \times {\bm
\lambda}|)} ,\label{hypVY1a}  \
\end{eqnarray}
which is valid when
\begin{eqnarray} &
\left\{
\begin{array}{l}
\quad 2 {\bm \rho}^2 - \sqrt{3}{\bm \rho} \cdot {\bm \lambda} \geq
- \rho \sqrt{{\bm \rho}^2 + 3 {\bm \lambda}^2 - 2 \sqrt{3}{\bm
\rho} \cdot {\bm \lambda}} \label{hypVY1aCond} \\
\quad 2 {\bm \rho}^2 + \sqrt{3}{\bm \rho} \cdot {\bm \lambda} \geq
- \rho \sqrt{{\bm \rho}^2 + 3 {\bm \lambda}^2 + 2 \sqrt{3}{\bm
\rho} \cdot {\bm \lambda}}  \\
\quad 3 {\bm \lambda}^2 - {\bm \rho}^2 \geq - \frac{1}{2} \sqrt{
({\bm \rho}^2 + 3 {\bm \lambda}^2)^{2} - 12 ({\bm \rho} \cdot {\bm
\lambda})^{2}}
\, .\\
\end{array} \right.
\end{eqnarray}
and three other angle-dependent two-body string, also called
V-string terms,  see Eqs. (\ref{hypVY1b})-(\ref{hypVY1d})
in Appendix \ref{a:Y_string}.

Due to the complexity of conditions Eq.\ (\ref{hypVY1aCond}) and Eqs. (\ref{hypVY1b})-(\ref{hypVY1d}),
and to the difficulties related to their implementation in calculations, there was a widespread lack
of use of the full Y-string potential Eq.\ (\ref{conf_Y}) in comparison to its dominant part,
the central Y-string potential $V_{\rm Y-central}$. 
In our hyperspherical harmonics approach, however, both the full Y-string potential and its central
part are treated in the same manner (just as the rest of the potentials) and present
no significant mathematical obstacles.
Both the central and the full Y-string potentials are decomposed into hyperspherical harmonics
and the resulting decomposition coefficients turn out close to each other, which renders
$V_{\rm Y-central}$ a good approximation to the full Y-string potential.

However, there is a physical reason that 
favors retaining only the central part of the Y-string potential over taking account of the
full potential: namely, the central Y-string potential $V_{\rm Y-central}$,
Eq. (\ref{hypVY1a}), has an exact dynamical O(2) symmetry, unlike the full potential,
Eq. (\ref{conf_Y}). To demonstrate this, we first show that the $V_{\rm Y-central}$
is a function of both Delves-Simonov hyper-angles $(\chi, \theta)$,
\begin{eqnarray}
V_{\rm Y-central}(R, \chi, \theta) &=& \sigma_{\rm Y} R \sqrt{\frac{3}{2}\left(1 + \sin 2 \chi
|\sin \theta| \right)} , \label{e:hypVY1a1} \
\end{eqnarray}
but a function of only one Smith-Iwai hyper-angle - the ``polar angle'' $\alpha$
\begin{eqnarray}
V_{\rm Y-central}(R, \alpha, \phi) &=& \sigma_{\rm Y} R \sqrt{\frac{3}{2}\left(1 + |\cos \alpha|
\right)}. \label{e:hypVY1b} \
\end{eqnarray}
This independence of the ``azimuthal'' Smith-Iwai hyper-angle $\phi$ means
that the associated component $\G$ of the hyper-angular momentum
(as in Ref. \cite{Salom:2017ekb}) is a
constant-of-the-motion. As this is actually a feature of the
$|{\bm \rho} \times {\bm \lambda}|$ term that is proportional
to the area of the triangle subtended by the three quarks,
the property is thus shared by all area-dependent potentials, such as the
central part of the Y-string Refs. \cite{Dmitrasinovic:2009ma}.

The expansion Eq. (\ref{e:Vdecomposition}) of the central Y-string potential
Eq. (\ref{e:hypVY1b}) in hyperspherical harmonics
\begin{eqnarray}
V_{\rm Y-central}(R, \alpha, \phi) &=& \sigma_{\rm Y} R \sqrt{\frac{3}{2}}
\sum_{K=0,4,...}^{\infty} v_{K0}^{\rm Y} {\cal Y}^{K0\Veig}_{00}(\alpha, \phi)
\label{e:hypVY2f} \\
&\equiv& V_{\rm eff.}^{\rm Y}(R)
\left(1 + \frac{v_{40}^{\rm Y}}{v_{00}^{\rm Y}} \pi \sqrt{{\pi}}
{\cal Y}^{40}_{000}(\alpha, \phi) + \ldots \right)  ,
\nonumber \
\end{eqnarray}
where $v_{K\G}^{\rm Y}$ are defined in Eq. (\ref{e:def_ME}),
runs over $O(6)$ hyper-spherical harmonics with $K = 0, 4, 8, \ldots$ and
zero value of the democracy quantum number $\G = 0$, as well, as vanishing angular
momentum $L = m = 0$ \footnote{We note that
the number $\G$ is sometimes also denoted as $G_3$ in the literature, \cite{Dmitrasinovic:2014}.}.
The numerical values are tabulated in Table \ref{tab:Potential_expansion}.

On the contrary, the expansion of the full Y-string potential Eq. (\ref{conf_Y})
has additional terms with ${\K = 0 ({\rm mod}~ 6),\G = 0 ({\rm mod}~ 6)}$, 
that spoil the dynamical O(2) symmetry of the potential in Eq. (\ref{hypVY1a}).
These terms are much smaller than the corresponding terms in the $\Delta$-string, QCD Coulomb
and logarithmic potentials, see Table \ref{tab:Potential_expansion}, and may therefore be neglected,
in leading approximation, with impunity. 
In Appendix \ref{a:Y_string} we illustrate how to evaluate the coefficient
$v_{\K = 6,\G =\pm6}^{\rm Y-string}$ and show its value in Table \ref{tab:Potential_expansion}.

\subsection{The QCD Coulomb potential}
\label{s:Coulomb_potential}

The QCD Coulomb potential Eq. (\ref{e:Coulomb}) is attractive in all three pairs, unlike the
electromagnetic one; 
in terms of Jacobi vectors it reads
\begin{equation}
\label{e:Coulomb}
V_{\rm Coulomb} = - \alpha_{\rm C} \sum_{i > j = 1}^3 |{\bf
x}_{i} - {\bf x}_{j}|^{-1}.
\end{equation}
\begin{eqnarray}
V_{\rm Coulomb} &=& - \alpha_{\rm C} \Bigg(\frac{1}{\sqrt{2 {\bm \rho}^2}} +
\frac{1}{\sqrt{\frac12 \left({\bm \rho}^2 + 3 {\bm \lambda}^2 - 2 \sqrt{3}{\bm
\rho} \cdot {\bm \lambda}\right)}} \nonumber \\
&+& \frac{1}{\sqrt{\frac12 \left({\bm \rho}^2 + 3 {\bm \lambda}^2 + 2 \sqrt{3}{\bm
\rho} \cdot {\bm \lambda}\right)}} \Bigg). \label{e:Coulomb2} \
\end{eqnarray}
The Coulomb potential's hyperspherical expansion is
\begin{eqnarray}
V_{\rm Coulomb}(R, \alpha, \phi) &=& V_{\rm Coulomb}(R) V_{\rm Coulomb}(\alpha, \phi) \label{e:Coulomb2a} \\
&=& V_{\rm Coulomb}(R) \sum_{K,\G}^{\infty} v_{K,\G}^{\rm Coulomb} {\cal Y}^{K \G \Veig}_{00}(\alpha, \phi)
 \nonumber \
\end{eqnarray}
where $V_{\rm Coulomb}(R) = - \alpha_{\rm C}/R$ and
the expansion coefficients $v_{K,\G}^{\rm Coulomb}$ are defined by the Coulomb analogon of
Eq. (\ref{e:def_ME})
and are tabulated in Table \ref{tab:Potential_expansion}.

We note that this and any other permutation symmetric sum of two-body potentials
(with the sole exception of the harmonic oscillator) has a
specific ``triple-periodic'' azimuthal $\phi$ hyper-angular dependence with the angular
period of $\frac23 \pi$. That provides additional selection rules for the ``democracy quantum number''
$\G$-dependent terms in this expansion, besides the $K=0,4,...$ rule for $\G = 0$ terms discussed above:
\begin{eqnarray}
\sum_{K \G}^{\infty} v_{K \G}^{\rm \Delta} {\cal Y}^{K \G \Veig}_{00}(\alpha, \phi)
&=&
\sum_{K=0,4,...}^{\infty} v_{K0}^{\rm \Delta} {\cal Y}^{K 0 \Veig}_{00}(\alpha, \phi) \nonumber \\
&+&
\sum_{K, \G =\pm 6}^{\infty} v_{K \G}^{\rm \Delta} {\cal Y}^{K \G \Veig}_{00}(\alpha, \phi) \nonumber \\
&+&
\sum_{K, \G =\pm 12}^{\infty} v_{K \G}^{\rm \Delta} {\cal Y}^{K \G \Veig}_{00}(\alpha, \phi) \nonumber \\
&+& ...
\label{e:hypDelta2d} \
\end{eqnarray}
Note that the values of all quantum numbers here are double of those in two spatial
dimensions (D=2), \cite{Dmitrasinovic:2014}. This has to do with the different integration
measures for D=2 and D=3 hyperspherical harmonics.

\subsection{The $\Delta$-string potential}
\label{s:Delta_potential}

The $\Delta$-string potential 
\begin{equation}
\label{conf_Delta}
V_{\Delta} = \sigma_{\Delta} \sum_{i > j = 1}^3 |{\bf
x}_{i} - {\bf x}_{j}|.
\end{equation}
written out in terms of Jacobi vectors reads
\begin{eqnarray}
V_{\Delta} &=& \sigma_{\Delta} \Bigg(\sqrt{2 {\bm \rho}^2} +
\sqrt{\frac12 \left({\bm \rho}^2 + 3 {\bm \lambda}^2 - 2 \sqrt{3}{\bm
\rho} \cdot {\bm \lambda}\right)} \nonumber \\
&+& \sqrt{\frac12 \left({\bm \rho}^2 + 3 {\bm \lambda}^2 +
2 \sqrt{3}{\bm \rho} \cdot {\bm \lambda}\right)} \Bigg).
\label{conf_Delta2} \
\end{eqnarray}



The $\Delta$-string potential Eq. (\ref{conf_Delta2}) in terms of Iwai-Smith angles
reads
\begin{eqnarray}
V_{\Delta}(R, \alpha, \phi) &=&
\sigma_{\Delta} R \Bigg(\sqrt{1 + \sin (\alpha) \sin \left(\frac{\pi }{6} - \phi\right)}
\nonumber \\
&+&  \sqrt{1 + \sin(\alpha) \sin \left(\phi + \frac{\pi}{6} \right)}
\nonumber \\
&+& \sqrt{1 - \sin(\alpha) \cos(\phi)}\Bigg)
\label{e:hypDelta1d} \
\end{eqnarray}
In order to find the general hyper-spherical harmonic expansion of the $\Delta$-string potential
we note that it factors into the hyper-radial $V_{\rm \Delta}(R)=\sigma_{\Delta} R$
and the hyper-angular part $V_{\rm \Delta}(\alpha, \phi)$
\begin{eqnarray}
V_{\rm \Delta}(R, \alpha, \phi) &=& V_{\rm \Delta}(R) V_{\rm \Delta}(\alpha, \phi) \nonumber \\
&=& V_{\rm \Delta}(R) \sum_{K,\G}^{\infty} v_{K,\G}^{\rm \Delta} {\cal Y}^{K \G \Veig}_{00}(\alpha, \phi)
\label{e:chypDelta1a} \
\end{eqnarray}
where the expansion coefficients $v_{K,\G}^{\Delta}$ are defined by the $\Delta$ analogon of
Eq. (\ref{e:def_ME})
and are tabulated in Table \ref{tab:Potential_expansion}.

\subsection{The general pair-wise power-law potential}
\label{s:power_potential}

Infinitely many permutation symmetric sums of two-body power-law potentials
have the generic form of Eq. (\ref{conf_Delta}) with different exponents $\epsilon$,
i.e.\ both the Coulomb and the $\Delta$-string potentials are two
special cases of the more general attractive homogeneous potential:
\begin{eqnarray}
V_{\epsilon}  &=& {\rm sgn(\epsilon)} \sigma_{\epsilon}
\sum_{i > j = 1}^3 |{\bf
x}_{i} - {\bf x}_{j}|^{\epsilon}
\nonumber \\
&=& {\rm sgn(\epsilon)} \sigma_{\epsilon} \Bigg(\Big(2 {\bm \rho}^2\Big)^{\epsilon / 2} +
\nonumber \\
&+&
\bigg(\frac12 \left({\bm \rho}^2 + 3 {\bm \lambda}^2 - 2 \sqrt{3}{\bm
\rho} \cdot {\bm \lambda}\right)\bigg)^{\epsilon / 2} \nonumber \\
&+& \bigg(\frac12 \left({\bm \rho}^2 + 3 {\bm \lambda}^2 +
2 \sqrt{3}{\bm \rho} \cdot {\bm \lambda}\right)\bigg)^{\epsilon / 2} \Bigg)
\label{power_epsilon_pot} \
\end{eqnarray}
where ${\rm sgn(\epsilon)} =  \epsilon/|\epsilon|$.
Note that in the special case of harmonic oscillator potential ($\epsilon=2$) the above form
degenerates into expression proportional to ${\bm \rho}^2 + {\bm \lambda}^2 = R^2$.

\begin{figure}[tbp]
\centerline{\includegraphics[width=3.25in,,keepaspectratio]{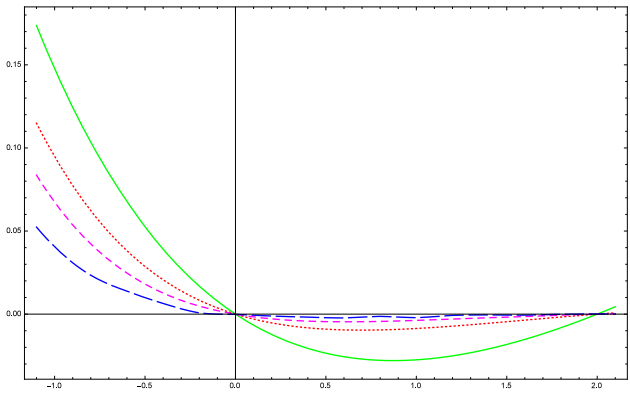}}
\caption{The graphs of the ratios $v_{4,0}^{\epsilon}/v_{00}^{\epsilon}$ (green, solid),
$v_{6,6}^{\epsilon}/v_{00}^{\epsilon}$ (red, dotted), $v_{8,0}^{\epsilon}/v_{00}^{\epsilon}$
(magenta, short dashes) and $v_{12,0}^{\epsilon}/v_{00}^{\epsilon}$ (blue, long dashes)
(listed in the decreasing order) as functions of the power $\epsilon$ in the potential
Eq. (\ref{power_epsilon_pot}).
One can see the tendency of the higher-order coefficients to diminish with increasing value of index $\K$.}
\label{f:PowerPotDecomposition}
\end{figure}

In Fig. 
\ref{f:PowerPotDecomposition}, 
we display the graphs of four ratios of h.s. expansion coefficients as functions of the exponent $\epsilon$.
There one can see that these coefficients depend smoothly on the exponent $\epsilon$ and that they
uniformly decrease with increasing value of index $\K$, in this class of potentials.
Numerical values of five expansion coefficients of potentials $V_{\rm Y}$, $V_{\Delta}$,
$V_{\rm Coulomb}$ and $V_{\rm Log}$ are shown in Table \ref{tab:Potential_expansion}.

\subsection{The Logarithmic potential}
\label{s:Log_potential}

The logarithmic potential 
\begin{equation}
\label{e:log}
V_{\rm Log} = \sigma_{\rm Log} \sum_{i > j = 1}^3 \log(|{\bf
x}_{i} - {\bf x}_{j}|).
\end{equation}
has a divergent short-distance and a steadily rising
long-distance part; thence it can be thought of as a linear combination of
the QCD Coulomb (with a homogeneity index $\alpha = -1$)
and a linear confining potential (with a homogeneity index $\alpha = 1$),
with a common homogeneity index equal to zero: $\alpha = 0$.
Note that this homogeneity condition boils down to an additive, rather
than multiplicative factorization of the potential:
\begin{eqnarray}
V_{\rm Log}(R, \alpha, \phi) &=& V_{\rm Log}(R) + V_{\rm Log}(\alpha, \phi) \nonumber \\
&=& V_{\rm Log}(R) + \sum_{K,\G}^{\infty} v_{K,\G}^{\rm Log} {\cal Y}^{K \G \Veig}_{00}(\alpha, \phi)
\label{e:Log2a} \
\end{eqnarray}

The logarithmic potential has been used with great success in the
heavy-quark-antiquark two-body problem: it reproduces the remarkable
mass-independence of the $c{\bar c} - J/\Psi$ and $b{\bar b} - \Upsilon$
spectra. It has not been used in the three-quark problem at all, 
to our knowledge.

\onecolumngrid
\begin{table}[tbh]
\begin{center}
\caption{Expansion coefficients $v_{KQ}$ of the Y- and $\Delta$-string as
well as of the Coulomb and Logarithmic
potentials in terms of O(6) hyper-spherical harmonics ${\cal Y}_{0,0}^{K,0,0}$,
for $K$ = 0, 4, 8, 12, respectively, and of the hyper-spherical harmonics
${\cal Y}_{0,0}^{6,\pm 6,0}$.}
\begin{tabular}{cccccc}
\hline \hline $(K,Q)$ & $v_{KQ}({\rm Y-central})$ & $v_{KQ}({\rm Y-string})$ &
$v_{KQ}(\Delta)$ & $v_{KQ}({\rm Coulomb})$ & $v_{KQ}(\rm Log)$ \\
\hline (0,0) & 8.18 & 8.22 & 16.04 & 20.04 
& -6.58 \\
\hline (4,0) & -0.443 & -0.398 & - 0.445 & 2.93 
& -1.21  \\
\hline (6,$\pm$6) & 0 & -0.027 & - 0.14 & 1.88
& -0.56  \\
\hline (8,0) & - 0.064 & -0.064 & - 0.04 & 1.41
& -0.33 \\
\hline (12,0) & -0.01 & -0.01 & 0 & 0 & -0.17 \\
\hline
\end{tabular}
\label{tab:Potential_expansion}
\end{center}
\end{table}
\twocolumngrid

\section{Results}
\label{s:Results}

In the following we present the $\K=0,\cdots,5$ shells' energy spectra, for two reasons:
(1) both as an example of the kind of results that one may expect as $\K$ increases,
and in order to settle some long-standing issues regarding the $\K=3$ shell,
\cite{Bowler:1981xh,Bowler:1982ck,Richard:1989ra};
(2) as an illustration of the methods, see Appendix \ref{a:calculations},
that were used in their calculation.
With regard to (2), we note that these examples are all purely algebraic, in the sense
that no numerical calculations were necessary, but that ceases to be the case as $\K$
increases beyond $\K > 8$, at first only for certain subsets of states, and ultimately,
for all states.

We note that we have already reported at a conference \cite{Salom:2016prag}, some of the
$\K = 4$ shell results, albeit without derivation.

\subsection{$\K=0,1,2$ shells}
\label{s:K2res}

\begin{figure}[tbp]
\centerline{\includegraphics[width=2.75in,,keepaspectratio]{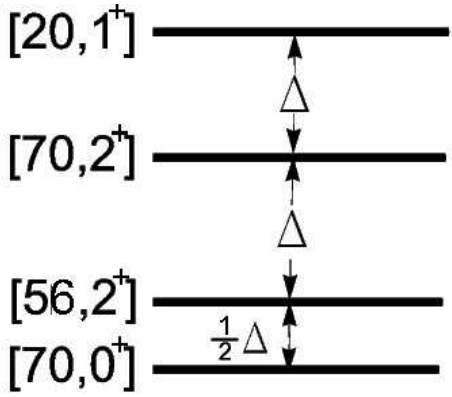}}
\caption{The K=2 spectrum of both the Y- and $\Delta$ strings in three dimensions.}
\label{f:K2spectrum}
\end{figure}

The $\K=0,1$ bands are affected only by the $v_{00}$ coefficient,
so, they need not be treated separately here,
whereas the $\K=2$ band is affected by the $v_{00}$ and $v_{40}$ coefficients.
The calculated energy splittings of $\K=2$ shell states depend only on the SU(6) multiplets,
\begin{eqnarray}
\left[ 20, 1^{+} \right] & \frac{1}{\pi \sqrt{\pi}} \Bigg( & v_{00} - \frac{1}{\sqrt{3}} v_{40} \Bigg) \nonumber \\
\left[ 70, 0^{+} \right] & \frac{1}{\pi \sqrt{\pi}} \Bigg( & v_{00} + \frac{1}{\sqrt{3}} v_{40}  \Bigg) \nonumber \\
\left[ 70, 2^{+} \right] & \frac{1}{\pi \sqrt{\pi}} \Bigg( & v_{00} - \frac{1}{5 \sqrt{3}} v_{40} \Bigg) \nonumber \\
\left[ 56, 2^{+} \right] & \frac{1}{\pi \sqrt{\pi}} \Bigg( & v_{00} + \frac{\sqrt{3}}{5} v_{40}  \Bigg)
\label{e:level_split_K2} , \
\end{eqnarray}
and the resulting spectrum is shown in Fig. \ref{f:K2spectrum}.
Our main concern is the energy splitting pattern among the states
within the K=2 hyper-spherical O(6) multiplet.
The hyper-radial matrix elements of the linear hyper-radial
potential are identical for all the (hyper-radial ground) states
in one K band. Therefore, as is well known, the energy differences among various
sub-states of a particular K band multiplet are integer multiples
of the energy splitting ``unit" $\Delta_{K} = \frac{1}{\pi \sqrt{\pi}}
\Bigg(\frac{1}{5 \sqrt{3}} - \frac{1}{\sqrt{3}} \Bigg) v_{40} =
- \frac{1}{\pi \sqrt{\pi}} \frac{4}{5 \sqrt{3}} v_{40}$.
Note, however, that this kind of spectrum is subject to the condition $v_{00} \neq 0$.

\subsection{$\K=3$ shell}
\label{s:K3res}

With only an area-dependent (i.e. $\phi$-independent)
the central Y-string potential $V_{\rm Y-central}$, Eq. (\ref{hypVY1a}),
in 3D, we find that the K=3 band $SU(6)$, or $S_3$ multiplets have one of four
possible energies shown in Eqs. (\ref{e:level_split_K3}) with
$v_{6\pm 6}^{\rm Y-central} = 0$.

\begin{figure}[tbp]
\centerline{\includegraphics[width=2.57in,,keepaspectratio]{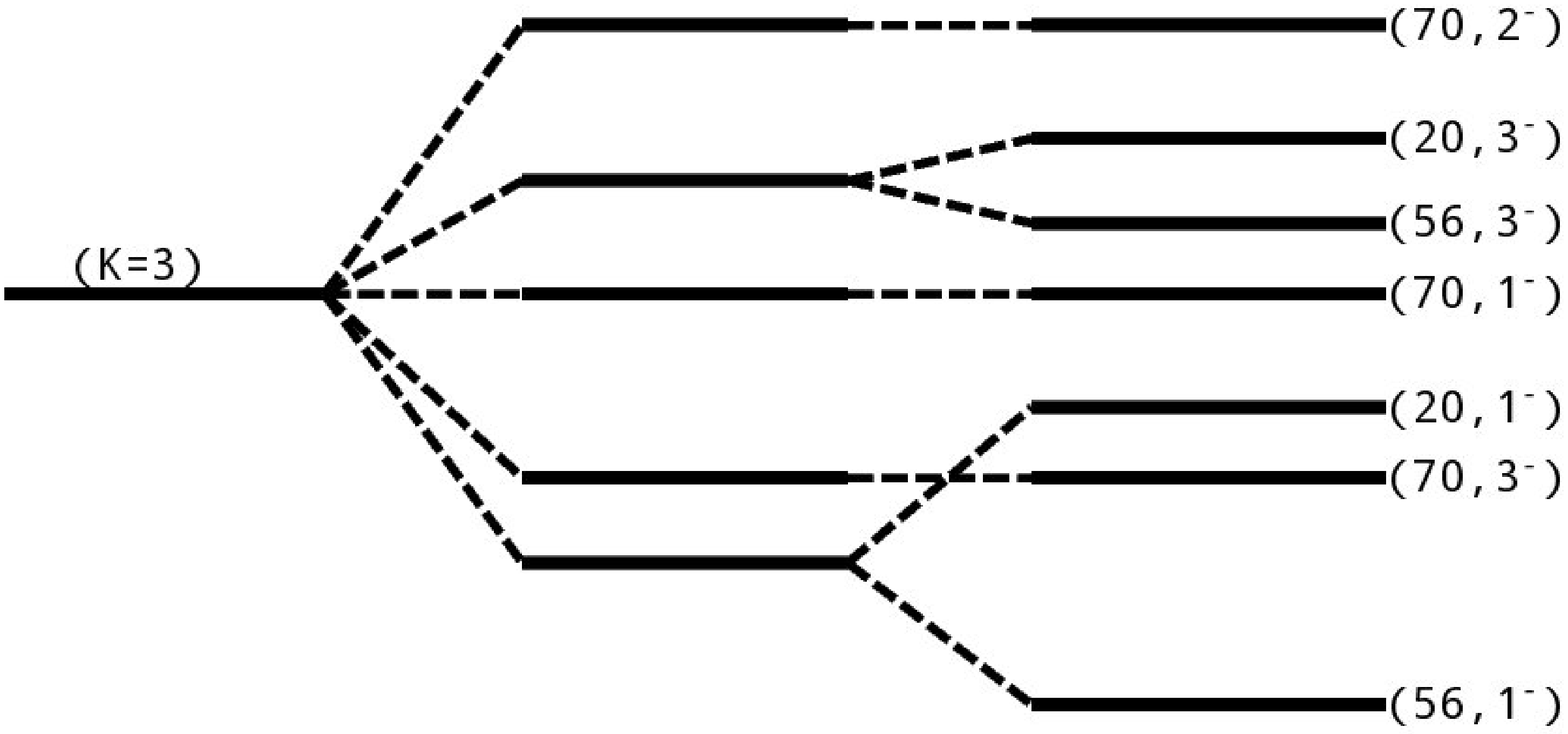}}
\caption{Schematic representation of the K=3 band in the energy spectrum
of the $\Delta$-string potential in three dimensions, following Ref. \cite{Richard:1989ra}.
The sizes of the two splittings (the $v_{40}^{\rm \Delta}$-induced
$\Delta$ and the subsequent $v_{6 \pm 6}^{\rm \Delta}$-induced splitting)
are not on the same scale, the latter having been increased, so as to be clearly visible.
The $\Delta$ here is the same as the $\Delta$ in the K=2 band.}
\label{f:K3spectrum}
\end{figure}

Upon introduction of the $\phi$-dependent two-body ``V-string'' potentials
$V_{\rm V-string}$, Eqs. (\ref{hypVY1b})-(\ref{hypVY1d}) into the full Y-string, the
$v_{6\pm 6}^{\rm Y-string}$ coefficient becomes $\neq 0$. 
After diagonalization of the $C_{[K'],[K]}$ matrix, one finds
further splittings among the previously degenerate states
$[70,1^-]$, $[56,3^-]$ and $[20,3^-]$,
as well as among the $[70,3^-]$, $[56,1^-]$ and $[20,1^-]$:
\begin{eqnarray}
\left[ 20, 1^{-} \right] & \frac{1}{\pi \sqrt{\pi}} \Bigg( &
v_{00} + \frac{1}{\sqrt{3}} v_{40} - \frac{2}{7} v_{66} \Bigg) \nonumber \\
\left[ 56, 1^{-} \right] & \frac{1}{\pi \sqrt{\pi}} \Bigg( &
v_{00} + \frac{1}{\sqrt{3}} v_{40} + \frac{2}{7} v_{66} \Bigg) \nonumber \\
\left[ 70,1^{-}\right] & \frac{1}{\pi \sqrt{\pi}} \Bigg( & v_{00} ~~~~~~~~~~~~~~~~~~~~~~~\Bigg)  \nonumber \\
\left[ 70,2^{-}\right] & \frac{1}{\pi \sqrt{\pi}} \Bigg( & v_{00} -\frac{1}{\sqrt{3}} v_{40} ~~~~~~~~~~\Bigg) \nonumber \\
\left[ 70,3^{-}\right] & \frac{1}{\pi \sqrt{\pi}} \Bigg( & v_{00} -\frac{1}{\sqrt{3}} v_{40} ~~~~~~~~~~\Bigg) \nonumber \\
\left[ 20,3^{-}\right] & \frac{1}{\pi \sqrt{\pi}} \Bigg( & v_{00} - \frac{\sqrt{3}}{7} v_{40}~ - ~~v_{66} \Bigg) \nonumber \\
\left[ 56,3^{-}\right] & \frac{1}{\pi \sqrt{\pi}} \Bigg( & v_{00} - \frac{\sqrt{3}}{7} v_{40}~ + ~~v_{66} \Bigg)
\label{e:level_split_K3} , \
\end{eqnarray}
where our $v_{66} < 0$ is negative, and Richard and Taxil's is positive.
These results are displayed in Fig. \ref{f:K3spectrum}.

For the $\K=3$ band in 3D, the energy splittings have been calculated by
Bowler and Tynemouth \cite{Bowler:1981xh}, \cite{Bowler:1982ck}
for two-body anharmonic potentials perturbing the harmonic oscillator and
confirmed and clarified by Richard and Taxil, Ref. \cite{Richard:1989ra},
in the hyper-spherical formalism with linear two-body potentials (the $\Delta$-string).

In hindsight, Richard and Taxil's  Ref. \cite{Richard:1989ra} separation of
$V_4(R)$ and $V_6(R)$ potentials' contributions is particularly
illuminating: the former corresponds precisely to our ``$\phi$-independent''
term $v_{40}$ and the latter to the ``$\phi$-dependent'' potential's
contribution to $v_{66}$.

As both the central Y-string and the $\Delta$-string contain the former, whereas
only the $\Delta$-string contains the latter, we see that the latter is
the source of different degeneracies/splittings in the spectra of these
two types of potentials.\footnote{This was not noted by Richard and Taxil,
Ref. \cite{Richard:1989ra}, however, so our contribution here is the (first)
demonstration of this fact in 3D case that has finally been confirmed in detail.}

\subsection{$\K=4$ shell}
\label{ss:K4}

\begin{figure}[tbp]
\centerline{\includegraphics[width=2.75in,,keepaspectratio]{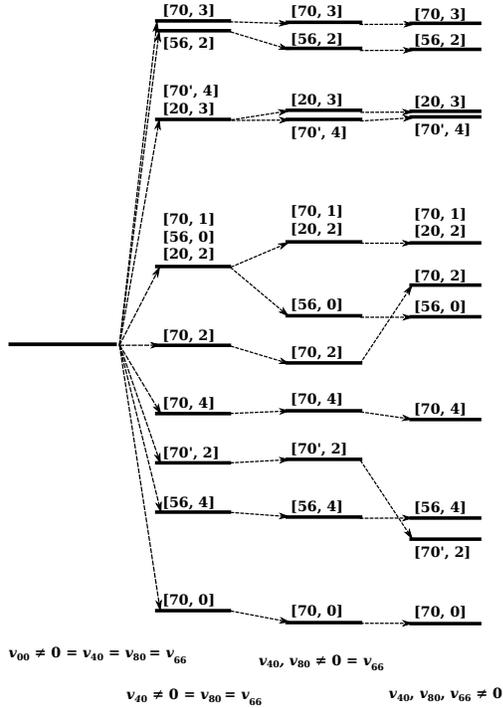}}
\caption{Schematic representation of the K=4 band in the energy spectrum
of three quarks in the $\Delta$-string potential.} 
\label{f:K4spectrum}
\end{figure}

In the $\K=4$ band $SU(6)$, or $S_3$ multiplets have one of the following 12 values of the
diagonalized $C$-matrix $C^\K_{\set{m_d}} \times \frac{v_{00}}{\pi \sqrt{\pi}}$,
from which one can evaluate the eigen-energies. We use the baryon-spectroscopic notation:
$[{\rm dim.} ,L^P]$, where ${\rm dim.}$ is the dimension of the  $SU_{\rm FS}(6)$ representation and
the correspondence with the representations of the permutation group $S_3$ is given as
$70 \leftrightarrow M$, $20 \leftrightarrow A$, $56 \leftrightarrow S$.
\begin{eqnarray}
&&  \left[ 70, 0^+ \right]: \frac{1}{\pi \sqrt{\pi}} \Bigg( v_{00} + \frac{\sqrt{3}}{2} v_{40}
+ \frac{1}{2\sqrt{5}} v_{80} \Bigg) \nonumber \\
&&\left[ 56, 0^{+}\right]: \frac{1}{\pi \sqrt{\pi}} \Bigg( v_{00} ~~~~~~~~~~~~~~
+ \frac{2}{\sqrt{5}} v_{80} \Bigg) \nonumber \\
&&\left[70,1^{+} \right]: \frac{1}{\pi \sqrt{\pi}} \Bigg( v_{00} ~~~~~~~~~~~~~~
- \frac{1}{\sqrt{5}} v_{80} \Bigg) \nonumber \\
&&\left[ 70, 2^+ \right]: \frac{1}{\pi \sqrt{\pi}}
\Bigg(v_{00} + \frac{1}{35} \Big(7 \sqrt{3} v_{4 0} + 2 \sqrt{5} v_{8 0} \nonumber \\
&&- 3 \sqrt{3 v_{4 0}^2 - 2 \sqrt{15} v_{4 0} v_{8 0} + 5 v_{8 0}^2 + 120 v_{6\pm 6}^2}\Big)
\Bigg) \nonumber \\
&&\left[ 70^{'}, 2^+ \right]: \frac{1}{\pi \sqrt{\pi}} \Bigg( v_{00}
+ \frac{1}{35} \Big(7 \sqrt{3} v_{4 0} + 2 \sqrt{5} v_{8 0} \nonumber \\
&&
+ 3 \sqrt{3 v_{4 0}^2 - 2 \sqrt{15} v_{4 0} v_{8 0} + 5 v_{8 0}^2 + 120 v_{6\pm 6}^2}\Big)
\Bigg) \nonumber \\
&&\left[ 56,2^+ \right]: \frac{1}{\pi \sqrt{\pi}}
\Bigg( v_{00} -\frac{12 \sqrt{3}}{35} v_{40} + \frac{ \sqrt{5}}{7} v_{80}
\Bigg) \nonumber \
\end{eqnarray}
\begin{eqnarray}
&&\left[20,2^+ \right]: \frac{1}{\pi \sqrt{\pi}}
\Bigg( v_{00} ~~~~~~~~~~~~~~~ -\frac{1}{\sqrt{5}} v_{80}
\Bigg) \nonumber \\
&& \left[ 20,3^{+}\right]: \frac{1}{\pi \sqrt{\pi}}
\Bigg( v_{00} - \frac{3 \sqrt{3}}{14} v_{40} - \frac{\sqrt{5}}{14} v_{80} \Bigg) \nonumber \\
&& \left[ 70,3^{+} \right]: \frac{1}{\pi \sqrt{\pi}}
\Bigg( v_{00} - \frac{5 \sqrt{3}}{14} v_{40} +
\frac{1}{14 \sqrt{5}} v_{80} \Bigg) \nonumber \\
&&\left[ 56, 4^+ \right]: \frac{1}{\pi \sqrt{\pi}} \Bigg( v_{00} +
\frac{5 \sqrt{3}}{14} v_{40} + \frac{3}{14 \sqrt{5}} v_{80} \Bigg) \nonumber \\
&&\left[ 70, 4^+ \right]: \frac{1}{\pi \sqrt{\pi}} \Bigg( v_{00} +
\frac{1}{42 \sqrt{5}} \Big(- 2 v_{8 0}  \nonumber \\
&& - \sqrt{1215 v_{4 0}^2 - 54 \sqrt{15} v_{4 0} v_{8 0} +
9 v_{8 0}^2 + 1280 v_{6\pm 6}^2} \Big) \Bigg) \nonumber \\
&&\left[ 70^{'}, 4^+ \right]: \frac{1}{\pi \sqrt{\pi}} \Bigg(
v_{00} + \frac{1}{42 \sqrt{5}} \Big(- 2 v_{8 0}  \nonumber \\
&&
+ \sqrt{1215 v_{4 0}^2 - 54 \sqrt{15} v_{4 0} v_{8 0} +
9 v_{8 0}^2 + 1280 v_{6\pm 6}^2} \Big)
\Bigg) 
\label{e:level_split_K4} . \
\end{eqnarray}
The $\Delta$-string results are shown in Fig. \ref{f:K4spectrum}.
Again, as the third coefficient $v_{6\pm 6}$ vanishes in the central
Y-string potential $V_{\rm Y-central}$ (which is without two-body terms),
or as it is roughly ten times smaller than usual, in the
full Y-string potential $V_{\rm Y-string}$,
the (second) observable difference between Y-string and
$\Delta$-string potentials shows up in the magnitude of splitting between the pairs of
$[ 70, 2^+ ], [ 70^{'}, 2^+ ]$ and $[ 70, 4^+ ],[ 70^{'}, 4^+ ] $ levels: the Y-string states
are ordered as shown in the third ($v_{6 \pm 6} = 0$) column in Fig. \ref{f:K4spectrum}.
 As explained earlier, the vanishing of $v_{6 \pm 6}$ follows from
the central Y-string potential's independence of the Iwai angle $\phi$, i.e., from
the dynamical ``kinematic rotations/democracy transformations'' O(2) symmetry,
\cite{Dmitrasinovic:2009ma,Dmitrasinovic:2014} associated with it.

Numerical results for other potentials are shown in Table \ref{tab:K4_3pot_numbers}.
Table \ref{tab:K4_3pot_numbers} shows that the
ordering of K = 4 states is not universally valid even for the (convex) potentials
considered here: note that although the three highest-lying multiplets
always come from the same set ($[70,3^+], [56,2^+], [20,3^+], [70^{'},4^+]$,
see Fig.  \ref{f:K4spectrum}), their orderings are different in these
potentials. That, of course, is a consequence of different ratios $v_{4 0}/v_{0 0}$,
$v_{6\pm 6}/v_{0 0}$ and $v_{8 0}/v_{0 0}$. This goes to show that
one cannot expect strongly restrictive ordering theorems to hold for
three-body systems, as they hold in the two-body problem, Ref. \cite{Grosse:1997xu}.
Nevertheless, even the present results are useful, as they indicate that
certain sets of multiplets are jointly lifted, or depressed, as a single group
in the spectrum, with ordering within the group being subject to the detailed
structure of the potential.

Of course, similar conclusions hold also for K = 3 spectrum splitting,
but are less pronounced, as that shell depends only on two numbers: the
ratios $v_{4 0}/v_{0 0}$ and $v_{6\pm 6}/v_{0 0}$. As the difference between
$\Delta$ and Y-string potentials is most pronounced in the value
of $v_{6\pm 6}$, that is the case where the distinction between these
two potentials is most clearly seen.
\begin{table}[tbh]
\begin{center}
\caption{The values of effective potentials in the 
Y-, $\Delta$-string, and (strong) Coulomb potentials
for various K = 4 states (for all allowed orbital waves L).}
\begin{tabular}{ccccc}
\hline \hline K
& $[SU(6),L^P]$ & $\langle \, {\rm V}_{\rm Y}/\sigma_{\rm Y} \,\rangle$
& $\langle \, {\rm V}_{\Delta}/\sigma_{\Delta} \,\rangle$
& $-\langle \, {\rm V}_{\rm C}/\alpha_{\rm C} \,\rangle$ \\
\hline
\hline
4 & $[56,0^+]$ & 1.45921 & 2.87122 & 3.82554 \\
4 & $[70,0^+]$ & 1.39729 & 2.80996 & 4.11043 \\
4 & $[70,1^+]$ & 1.47483 & 2.88587 & 3.48449 \\
4 & $[56,2^+]$ & 1.51372 & 2.92453 & 3.36709 \\
4 & $[20,2^+]$ & 1.47483 & 2.88587 & 3.48449 \\
4 & $[70,2^+]$ & 1.44997 & 2.87749 & 4.13184 \\
4 & $[70^{'},2^+]$ & 1.43052 & 2.82683 & 3.49379 \\
4 & $[70,3^+]$ & 1.51906 & 2.92963 & 3.281 \\
4 & $[20,3^+]$ & 1.50137 & 2.91213 & 3.36239 \\
4 & $[56,4^+]$ & 1.4187 & 2.83095 & 3.94783 \\
4 & $[70,4^+]$ & 1.44036 & 2.85066 & 3.3656 \\
4 & $[70^{'},4^+]$ & 1.49938 & 2.91178 & 3.81992 \\
\hline
\end{tabular}
\label{tab:K4_3pot_numbers}
\end{center}
\end{table}

On the phenomenological side, some eigen-energies of three quarks in the $\K=4$ shell
have been calculated in Ref. \cite{Stassart:1997vk} using a variational method based
on harmonic oscillator wave functions. These calculations
included the $\Delta$-string, Y-string and Coulomb potentials, all at once, as well
as a relativistic kinetic energy (this kinetic energy violates the O(6) symmetry).
Each one of these three terms in the potential is homogenous, but their sum is not -
therefore the individual contributions of these terms to the total/potential energy
cannot be compared directly with the results of their separate calculations.
Moreover, each term in the Hamiltonian breaks the O(6) symmetry differently,
thus inducing different splittings of energy spectra.
These facts prevent us from directly comparing our results with Ref. \cite{Stassart:1997vk},
but the overall trend for groups of states seem to be in agreement with our results, see
Table \ref{tab:K4_3pot_numbers}, for comparison.

\subsection{$\K=5$ shell}
\label{ss:K5}


With a 
$\phi$-independent central Y-string potential in 3D, we find
that the $\K=5$ band $SU(6)$, or $S_3$ multiplets have one of 15 
different energies. Upon introduction of a $\phi$-dependent (``two-body")
component of the potential, proportional to $v_{6 6}$,
and upon diagonalization of the $C_{[K'],[K]}$ matrix, one finds four new
splittings between previously degenerate states
1) $[56, 2^{-}], [20,2^{-}]$,
2) $[56^{'}, 4^-], [20^{'}, 4^-]$,
3) $[70~, 1^-], [70^{'}, 1^- ]$,
4) $[70~, 5^-], [70^{'}, 5^- ]$,
as well as three non-degenerate states whose energies are shifted by $v_{6 6}$.
These algebraic results are summarized in
\begin{eqnarray}
\left[ 70, 1^- \right] & \frac{1}{\pi \sqrt{\pi}} \Bigg( & v_{00} +
\frac{5 \sqrt{3} v_{4 0} + 3 \sqrt{5} v_{8 0}}{20}  \nonumber \\
& - & \frac{\sqrt{75 v_{4 0}^2 - 10 \sqrt{15} v_{4 0} v_{8 0}
+ 5 v_{8 0}^2 + 96 v_{6\pm 6}^2} }{20} \Bigg) 
\nonumber \\
\left[ 70^{'}, 1^- \right] & \frac{1}{\pi \sqrt{\pi}} \Bigg( & v_{00} +
\frac{5 \sqrt{3} v_{4 0} + 3 \sqrt{5} v_{8 0}}{20} \nonumber \\
& + & \frac{\sqrt{75 v_{4 0}^2 - 10 \sqrt{15} v_{4 0} v_{8 0}
+ 5 v_{8 0}^2 + 96 v_{6\pm 6}^2} }{20} \Bigg) \nonumber \\ 
\left[ 56, 1^- \right] & \frac{1}{\pi \sqrt{\pi}} \Bigg( & v_{00} + \frac{1}{2 \sqrt{3}} v_{40} -\frac{2}{5}v_{66}
- \frac{1}{2 \sqrt{5}} v_{80} \Bigg) \nonumber \\ 
\left[ 20, 1^- \right] & \frac{1}{\pi \sqrt{\pi}} \Bigg( & v_{00} + \frac{1}{2 \sqrt{3}} v_{40} +\frac{2}{5}v_{66}
- \frac{1}{2 \sqrt{5}} v_{80} \Bigg) \nonumber \\ 
\left[ 70, 2^- \right] & \frac{1}{\pi \sqrt{\pi}} \Bigg( & v_{00} - \frac{1}{2 \sqrt{3}} v_{40} ~~~~~~~~~~
- \frac{1}{2 \sqrt{5}} v_{80} \Bigg) \nonumber \\ 
\left[ 56, 2^- \right] & \frac{1}{\pi \sqrt{\pi}} \Bigg( & v_{00} ~~~~~~~~~~~~~~
+ \frac{3}{5} v_{66} - \frac{1}{\sqrt{5}} v_{80} \Bigg) \nonumber \\ 
\left[ 20, 2^- \right] & \frac{1}{\pi \sqrt{\pi}} \Bigg( & v_{00} ~~~~~~~~~~~~~~
- \frac{3}{5} v_{66} - \frac{1}{\sqrt{5}} v_{80}  \Bigg) \nonumber \ 
\end{eqnarray}
\begin{eqnarray}
\left[ 56, 3^- \right] & \frac{1}{\pi \sqrt{\pi}} \Bigg( & v_{00} + \frac{1}{2 \sqrt{3}} v_{40} +
\frac{14}{15}v_{66} + \frac{7}{6 \sqrt{5}} v_{80} \Bigg) \nonumber \\ 
\left[ 20, 3^- \right] & \frac{1}{\pi \sqrt{\pi}} \Bigg( & v_{00} + \frac{1}{2 \sqrt{3}} v_{40} -
\frac{14}{15}v_{66} + \frac{7}{6 \sqrt{5}} v_{80}  \Bigg) \nonumber \\ 
\left[ 70, 4^- \right] & \frac{1}{\pi \sqrt{\pi}} \Bigg( & v_{00} - \frac{1}{2 \sqrt{3}} v_{40}
~~~~~~~~~~~~  - \frac{1}{2 \sqrt{5}} v_{80} \Bigg) \nonumber \\ 
\left[ 56, 4^- \right] & \frac{1}{\pi \sqrt{\pi}} \Bigg( & v_{00} - \frac{7}{6 \sqrt{3}} v_{40}
- \frac{2}{15} v_{66} + \frac{1}{6 \sqrt{5}} v_{80} \Bigg) \nonumber \\ 
\left[ 20, 4^- \right] & \frac{1}{\pi \sqrt{\pi}} \Bigg( & v_{00} - \frac{7}{6 \sqrt{3}} v_{40}
+ \frac{2}{15} v_{66} + \frac{1}{6 \sqrt{5}} v_{80} \Bigg) \nonumber \\ 
\left[ 70, 5^{-} \right] & \frac{1}{\pi \sqrt{\pi}} \Bigg(& v_{00} +
\frac{\sqrt{3}}{18} v_{4 0} + \frac{\sqrt{5}}{30} v_{8 0} \nonumber \\
+  & \frac{\sqrt{5}}{165} &
\sqrt{ 1815 v_{4 0}^2 + 66 \sqrt{15} v_{4 0} v_{8 0} + 9 v_{8 0}^2 + 968 v_{6\pm 6}^2} \Bigg)
\nonumber \\ 
\left[70^{'}, 5^- \right] & \frac{1}{\pi \sqrt{\pi}} \Bigg(& v_{00} +
\frac{\sqrt{3}}{18} v_{4 0} + \frac{\sqrt{5}}{30} v_{8 0} \nonumber \\
& - \frac{\sqrt{5}}{165} & \sqrt{1815 v_{4 0}^2 + 66 \sqrt{15} v_{4 0} v_{8 0} + 9 v_{8 0}^2 + 968 v_{6\pm 6}^2}\Bigg) 
\nonumber \\
\left[ 56 , 5^- \right] & \frac{1}{\pi \sqrt{\pi}} \Bigg(& v_{00} + \frac{1}{2 \sqrt{3}} v_{40} +
\frac{8}{15} v_{66} - \frac{19}{66 \sqrt{5}} v_{80} \Bigg) \nonumber \\ 
\left[ 20, 5^- \right] & \frac{1}{\pi \sqrt{\pi}} \Bigg(& v_{00} + \frac{1}{2 \sqrt{3}} v_{40}
- \frac{8}{15} v_{66} - \frac{19}{66 \sqrt{5}} v_{80} \Bigg) 
\label{e:level_split_K5} \
\end{eqnarray}

\begin{figure}[tbp]
\centerline{\includegraphics[width=3.5in,,keepaspectratio]{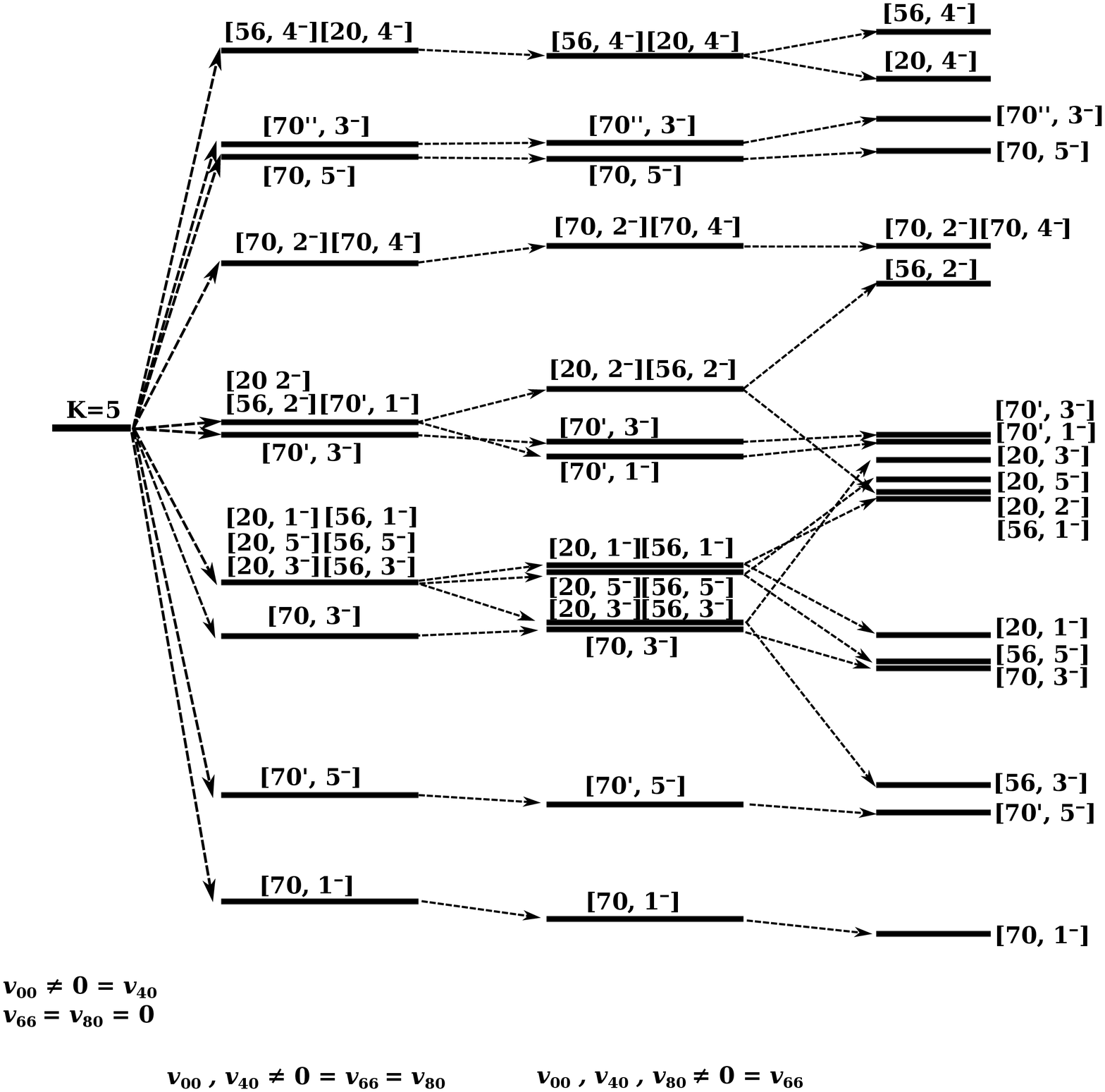}}
\caption{Schematic representation of the K=5 band in the energy spectrum
of the $\Delta$- and Y-string potentials in three dimensions.
The sizes of the two splittings (the $v_{40}^{\rm \Delta}$-induced
$\Delta$ and the subsequent $v_{80}^{\rm \Delta}$-induced splitting)
are not on the same scale.
The $\Delta$ here is not the same as the $\Delta$ in the K=2 band.}
\label{f:K5spectrum}
\end{figure}

The numerical results are displayed in Table \ref{tab:K5_3pot_numbers} for different potentials,
and in Table \ref{tab:K5_Y_splitting} for the $\Delta$-string potential it has been
broken up into contributions of different multipoles of the potential
and are graphically displayed in Fig. \ref{f:K5spectrum}.
\begin{table}[tbh]
\begin{center}
\caption{The values of the effective potential matrix elements
for the Y-, $\Delta$-string, and (strong) Coulomb potential,
and various K = 5 states (for all allowed orbital waves L).}
\begin{tabular}{ccccc}
\hline \hline K
& $[SU(6),L^P]$ & $\langle \, {\rm V}_{\rm Y}/\sigma_{\rm Y} \,\rangle$
& $\langle \, {\rm V}_{\Delta}/\sigma_{\Delta} \,\rangle$
& $-\langle \, {\rm V}_{\rm C}/\alpha_{\rm C} \,\rangle$ \\
\hline
\hline
5 & $[70,1^-]$ & 1.39729 & 2.80778 & 2.55667 \\
5 & $[70^{'},1^-]$ & 1.46442 & 2.87829 & 2.85542 \\
5 & $[56,1^-]$ & 1.44898 & 2.87055 & 2.46858 \\
5 & $[20,1^-]$ & 1.44898 & 2.85059 & 2.5953 \\
5 & $[70,2^-]$ & 1.49547 & 2.90629 & 2.32887 \\
5 & $[20,2^-]$ & 1.47483 & 2.87091 & 2.47611 \\
5 & $[56,2^-]$ & 1.47483 & 2.90084 & 2.28602 \\
5 & $[70,3^-]$ & 1.46682 & 2.84167 
& 2.41462 \\ 
5 & $[70^{'},3^-]$ & 1.44037 & 2.8887 
& 2.30016 \\ 
5 & $[70^{''},3^-]$ & 1.5103 & 2.92104 
& 2.67414 \\ 
5 & $[56,3^-]$ & 1.44031 & 2.82915 & 2.84424 \\
5 & $[20,3^-]$ & 1.44031 & 2.87571 & 2.54855 \\
5 & $[70,4^-]$ & 1.49547 & 2.90629 & 2.32887 \\
5 & $[56,4^-]$ & 1.52299 & 2.93685 & 2.23815 \\
5 & $[20,4^-]$ & 1.52299 & 2.93020 & 2.28039 \\
5 & $[70,5^-]$ & 1.50797 & 2.91991 & 2.75234 \\
5 & $[70^{'},5^-]$ & 1.41405 & 2.82520 & 2.30772 \\
5 & $[56,5^-]$ & 1.44788 & 2.84623 & 2.63735 \\
5 & $[20,5^-]$ & 1.44788 & 2.87283 & 2.46839 \\
\hline
\end{tabular}
\label{tab:K5_3pot_numbers}
\end{center}
\end{table}
\begin{table}[tbh]
\begin{center}
\caption{The values of the effective three-body $\Delta$-string potential
divided by the string tension $\sigma_{\Delta}$,
$\langle \, {\rm V}_{\Delta}(v_{0,0}, v_{4,0}, v_{6,6}, v_{8,0})/\sigma_{\Delta} \,\rangle$
as a function of the expansion coefficients $(v_{0,0}, v_{4,0}, v_{6,6}, v_{8,0})$, for various
$\K = 5$ states (for all allowed orbital waves $L$). Here
$\langle \, {\rm V}_{\Delta}(A)/\sigma_{\Delta}  \,\rangle = \langle \, {\rm V}_{\Delta}(v_{0,0}, v_{4,0} \neq 0 =
v_{6,6} = v_{8,0})/\sigma_{\Delta}  \,\rangle$
and $\langle \, {\rm V}_{\Delta}(B)/\sigma_{\Delta}  \,\rangle =
\langle \, {\rm V}_{\Delta}(v_{0,0}, v_{4,0} , v_{8,0} \neq 0 =
v_{6,6})/\sigma_{\Delta}  \,\rangle$}
\begin{tabular}{ccccc}
\hline \hline K
& $[SU(6),L^P]$ & $\langle \, {\rm V}_{\Delta}(A)/\sigma_{\Delta}  \,\rangle$
& $\langle \, {\rm V}_{\Delta}(B)/\sigma_{\Delta}  \,\rangle$
& $\langle \, {\rm V}_{\Delta}/\sigma_{\Delta} \,\rangle$ \\
\hline
\hline
5 & $[70,1^-]$ & 2.8124 & 2.80996 & 2.80778 \\
5 & $[70^{'},1^-]$ & 2.88099 & 2.87611 & 2.87829 \\
5 & $[56,1^-]$ & 2.85813 & 2.86057 & 2.87055 \\
5 & $[20,1^-]$ & 2.85813 & 2.86057 & 2.85059 \\
5 & $[70,2^-]$ & 2.90385 & 2.90629 & 2.90629 \\
5 & $[56,2^-]$ & 2.88099 & 2.88587 & 2.87091 \\
5 & $[20,2^-]$ & 2.88099 & 2.88587 & 2.90084 \\
5 & $[70,3^-]$ & 2.85051 & 2.85214 & 2.84167 \\ 
5 & $[70^{'},3^-]$ & 2.87918 & 2.87827 & 2.8887 \\ 
5 & $[70^{''},3^-]$ & 2.92091 & 2.921 & 2.92104 \\ 
5 & $[56,3^-]$ & 2.85813 & 2.85243 & 2.82915 \\
5 & $[20,3^-]$ & 2.85813 & 2.85243 & 2.87571 \\
5 & $[70,4^-]$ & 2.90385 & 2.90629 & 2.90629 \\
5 & $[56,4^-]$ & 2.93434 & 2.93352 & 2.93685 \\
5 & $[20,4^-]$ & 2.93434 & 2.93352 & 2.93020 \\
5 & $[70,5^-]$ & 2.91909 & 2.91872 & 2.91991 \\
5 & $[70^{'},5^-]$ & 2.82764 & 2.82639 & 2.82520 \\
5 & $[56,5^-]$ & 2.85813 & 2.85953 & 2.84623 \\
5 & $[20,5^-]$ & 2.85813 & 2.85953 & 2.87283 \\
\hline
\end{tabular}
\label{tab:K5_Y_splitting}
\end{center}
\end{table}

\section{Discussion and comparison with previous calculations}
\label{s:Discussion}

1) The present results are meant (only) as examples of what can be done: these calculations
can be extended with $\K$ increasing {\it ad infinitum}, with the help of O(6) matrix elements
that are functions
of O(6) Clebsch-Gordan coefficients, which can be found in Ref. \cite{Salom:2017ekb}. This
is subject to the proviso that at some value of $\K$ the calculations must necessarily
become numerical.

2) The algebraic results shown in Sect. \ref{s:Results} do not hold for the QCD Coulomb potential,
as the QCD Coulomb hyper-radial potential $- \alpha_C/R$ Eq. (\ref{e:Coulomb2a})
has a dynamical O(7) symmetry and therefore accidental degeneracies are expected to appear.
That symmetry is broken by the hyper-angular part of the Coulomb three-body potential
in a manner that still remains to be explored.


3) In the $\K= 2$ band/shell of the three-body energy spectrum the eigen-energies depend
on two  coefficients $(v_{00}, v_{40})$, and the splittings among various levels depend
only on the (generally small) ratio $v_{40}/v_{00}$. This means that the eigen-energies
form a fixed pattern (``ordering'') that does not depend on the shape of the three-body
potential. The actual size of the $\K= 2$ shell energy splitting depends on the small
parameter $v_{40}/v_{00}$, provided that the potential is permutation symmetric.
This fact was noticed almost 40 years ago, Refs. \cite{Gromes:1976cr,Isgur:1978wd},
and it suggested that similar patterns might exist in higher-$\K$ shells.

The practical advantage of permutation-adapted hyperspherical harmonics over the conventional ones
is perhaps best illustrated here: the $\K = 2$ shell splittings in the Y- and $\Delta$-string
potentials were obtained, after some complicated calculations using conventional hyperspherical
harmonics in Ref. \cite{Dmitrasinovic:2009dy},
whereas here they follow from the calculation of four (simple) hyper-angular matrix elements.

4) Historically, extensions of this kind of calculations to higher ($\K \geq 3$) bands,
for general three-body potentials turned out more problematic than expected:
Bowler et. al, Ref. \cite{Bowler:1981xh}, published a set of predictions for the $\K=3, 4$ bands,
which were later questioned by Richard and Taxil's $\K=3$ hyperspherical harmonic calculation,
Ref. \cite{Richard:1989ra}; see also Refs. \cite{Stancu:1991cz,Stassart:1997vk}.
This controversy had not been resolved up to the present day, to our knowledge, so we address that
problem first: In the $\K=3$ case the energies depend on three coefficients $(v_{00}, v_{40},
v_{6\pm 6})$, and there is no mixing of multiplets, so all eigen-energies can be expressed in
simple closed form that agrees with Ref. \cite{Richard:1989ra} and depends on two small parameters
$v_{40}/v_{00}, v_{6\pm 6}/v_{00}$.

Note that the coefficient $v_{6\pm 6}$ vanishes in the (simplified) central Y-string potential
(without two-body terms)
and thus causes the first potentially observable difference betwen Y- and $\Delta$-string potentials:
the splittings between $\left[20, 1^- \right]$, and $\left[56, 1^- \right]$, as well as between
$\left[20, 3^- \right]$, and $\left[56, 3^- \right]$. 
The actual value of $v_{6\pm 6}$ in the exact Y-string potential is so small, as to be negligible
compared with the other two coefficients, $v_{00}, v_{40}$, in its expansion.

5) Note that from Eq. (7) it follows that there must exist an upper limit on the values of the ratios
$|v_{40}/v_{00}| \leq \sqrt{3}$, and from Eq. (8) it follows $|v_{6\pm 6}/v_{00}| \leq 7/2$.
If these limits are exceeded, the overall sign of the effective potential flips, and the
solution (motion) becomes unbound. This example clearly shows the limitations of the
present method. However, the physically interesting potentials considered in Sect. \ref{s:potential}
all satisfy inequalities $v_{00} \gg |v_{40}|$, and $v_{00} \gg |v_{6\pm 6}|$, as can be seen
in Table \ref{tab:Potential_expansion} and Fig. \ref{f:PowerPotDecomposition},
which shows that this method may be applied here.

6) The above points (2) and (3) display possible ``fault line(s)'' in the predictions of ordering
of shells with different values of $\K$: in case $v_{00} = 0$, the $\K=0,1$ shells become unbound
to leading (adiabatic) order, and their binding becomes a question of higher-order (non-adiabatic) effects.

7) We shall not attempt a numerical predictions of triple-heavy hyperon masses here, for the following reasons:
(a) the mass of heavy quark(s) $m_Q$ is not precisely known in the three-quark environment;
(b) the QCD coupling constant $\alpha_S$ is not known in this environment;
(c) the value of the effective string tension $\sigma$ is not known in this environment;
(d) the spin-dependent interactions, which are not included here, may significantly influence the results.
Nevertheless, nothing prevents the interested reader from inserting his/her favourite values
of $m_Q$, $\alpha_S$ and $\sigma$ into our formulae to obtain some definite predictions.

8) There are several possible straightforward extensions of the present work: (a) to equal mass
systems with a relativistic kinematic energy; (b) to two identical and one distinct quark
systems. Both extensions break the O(6) symmetry further still, but can be treated within
the present approach, with certain {\it caveats}.

9) Note that we have kept the full $SU_F(3)$, $SU_{FS}(6)$ notation for the three-quark
states, even though there can be only one flavour, with three identical heavy quarks.
This is in order to keep maximum generality, and to allow potential future extension to
relativistic light quark systems (c.f. \cite{Stancu:1991cz,Stassart:1997vk}).

10) The present formalism allows a (mathematically proper) extension of the
Regge theory/trajectories \cite{Regge:1959} to three-quark systems,
as well as an extension of Birman-Schwinger's results \cite{Birman:1961,Schwinger:1961}
about the number of bound states of a Schr\" odinger equation in a given potential.

11) The present formalism allows an extension to atomic and molecular physics, as well,
albeit with significant modifications: (a) atomic systems are subject to Coulomb
potential, which leads to a higher dynamical symmetry, that needs to be taken into account;
and (b) molecular systems are bound by inhomogenous potentials, such as the Lennard-Jones
one, which must be treated differently.

\section{Summary and Conclusions}
\label{s:Summary}

In summary, we have reduced the non-relativistic (quantum) three-identical-body problem
to a single ordinary differential equation for the hyper-radial wave function
with coefficients multiplying the homogenous hyper-radial potential that are determined
entirely by O(6) group-theoretical arguments, see \cite{Salom:2016vcw,Salom:2017ekb}.
That equation can be solved in the same way as the radial Schr{\" o}dinger
equation in 3D. The breaking of the O(6) symmetry by the three-quark potential
determines the ordering of states within different shells in the energy spectrum.

The dynamical $O(2)$ symmetry of the Y-string potential was discovered in
Ref. \cite{Dmitrasinovic:2009ma}, with the permutation group $S_3 \subset O(2)$ as
the subgroup of the dynamical $O(2)$ symmetry. The existence of an additional dynamical
symmetry strongly suggested an algebraic approach, such as that used
in two-dimensional space, in Ref. \cite{Dmitrasinovic:2014}.
In three dimensions (3D) the hyper-spherical symmetry group is O(6),
and the residual dynamical symmetry of the potential is
$S_3 \otimes SO(3)_{rot} \subset O(2) \otimes SO(3)_{rot}
\subset O(6)$, where $SO(3)_{rot}$ is the rotational symmetry associated
with the (total orbital) angular momentum $L$.
We showed how the energy eigenvalues
can be calculated as functions of the three-body potential's (hyper-)spherical
harmonics expansion coefficients $v_{\K,\G}^{\rm 3-body}$ and O(6) Clebsch-Gordan
coefficients that are evaluated in Ref. \cite{Salom:2017ekb}.

We have used these results to calculate the energy splittings of various states
(or $SU_{FS}(6)$ and $S_3$ multiplets) in the $\K \leq 5$ shells of the Y-,
$\Delta$-string and Coulomb potential spectra.
The ordering of bound states has its most immediate application
in the physics of three confined quarks, where the question was raised originally,
Refs. \cite{Gromes:1976cr,Isgur:1978wd,Bowler:1981xh,Richard:1989ra}.
We have shown that in the $\K \geq 3$ shells a clear difference appears between
the spectra of the Y- and $\Delta$-string models of confinement.
That is also the first explicit consequence of the dynamical O(2) symmetry of
the ``Y-string'' potential.

We stress the algebraic nature of our results,
as this method can be used to obtain predictions for arbitrarily large $\K$ values,
which calculations must necessarily be numerical, however, as soon as
the number of states that are mixed exceeds five.

The results presented here do not represent the outer boundaries of applicability
of our method, but are rather just illustrative examples, with a view to
its application to atomic, molecular and nuclear physics.

\acknowledgments
We thank Mr. Aleksandar Bojarov for drawing the Figures 2 - 5.
This work was financed by the Serbian Ministry of Science and
Technological Development under grant numbers OI 171031, OI 171037 and III 41011.

\appendix

\section{Evaluation of obtuse-angled two-body contributions to the Y-string}
\label{a:Y_string}

As stated in Sect. \ref{s:HS3D} at obtuse angles ($\geq 120^{0}$) there are two-body contributions
to the Y-string potential that break the dynamical O(2) symmetry of Eq. (\ref{e:hypVY1b}).
Therefore, the expansion coefficient $v_{\K = 6,\G = \pm 6}^{\rm Y-string}$ of
the full potential is not zero $v_{6, \pm 6}^{\rm Y-string} \neq 0$.



Three angle-dependent two-body string in terms of Jacobi vectors ${\bm \rho}, {\bm \lambda}$ are,
see Ref. \cite{Dmitrasinovic:2009dy},
\onecolumngrid
\begin{subequations}
\begin{eqnarray}
V_{\rm V-string} &=& \sigma  \left( \sqrt{\frac{1}{2}({\bm \rho}^2 +
3 {\bm \lambda}^2 + 2 \sqrt{3}{\bm \rho} \cdot {\bm \lambda})} +
\sqrt{\frac{1}{2}({\bm \rho}^2 + 3 {\bm \lambda}^2 - 2
\sqrt{3}{\bm \rho} \cdot {\bm \lambda})}\right) \label{hypVY1b} \\
&& \textrm{when}  \left\{
\begin{array}{l}
\quad 2 {\bm \rho}^2 - \sqrt{3}{\bm \rho} \cdot {\bm \lambda} \geq
- \rho \sqrt{{\bm \rho}^2 + 3 {\bm \lambda}^2 - 2 \sqrt{3}{\bm \rho}
\cdot {\bm \lambda}} \nonumber \\
\quad 2 {\bm \rho}^2 + \sqrt{3}{\bm \rho} \cdot {\bm \lambda} \geq
- \rho \sqrt{{\bm \rho}^2 + 3 {\bm \lambda}^2 + 2 \sqrt{3}{\bm
\rho} \cdot {\bm \lambda}}  \\
\quad 3 {\bm \lambda}^2 - {\bm \rho}^2 \leq - \frac{1}{2} \sqrt{
({\bm \rho}^2 + 3 {\bm \lambda}^2)^{2} -
12 ({\bm \rho} \cdot {\bm \lambda})^{2}}\\
\end{array} \right. \\
V_{\rm V-string} &=& \sigma \left( \sqrt{2} \rho +
\sqrt{\frac{1}{2}({\bm \rho}^2 + 3 {\bm \lambda}^2 + 2
\sqrt{3}{\bm \rho} \cdot {\bm \lambda})} \right)
\label{hypVY1c} \\
&& \textrm{when}  \left\{
\begin{array}{l}
\quad 2 {\bm \rho}^2 - \sqrt{3}{\bm \rho} \cdot {\bm \lambda} \geq
- \rho \sqrt{{\bm \rho}^2 + 3 {\bm \lambda}^2 - 2 \sqrt{3}{\bm
\rho} \cdot {\bm \lambda}} \nonumber \\
\quad 2 {\bm \rho}^2 + \sqrt{3}{\bm \rho} \cdot {\bm \lambda} \leq
- \rho \sqrt{{\bm \rho}^2 + 3 {\bm \lambda}^2 + 2 \sqrt{3}{\bm
\rho} \cdot {\bm \lambda}} \nonumber \\
\quad 3 {\bm \lambda}^2 - {\bm \rho}^2 \geq - \frac{1}{2} \sqrt{
({\bm \rho}^2 + 3 {\bm \lambda}^2)^{2} - 12 ({\bm \rho} \cdot {\bm
\lambda})^{2}}
\end{array} \right. \\
V_{\rm V-string} &=& \sigma \left( \sqrt{2} \rho +
\sqrt{\frac{1}{2}({\bm \rho}^2 + 3 {\bm \lambda}^2 - 2
\sqrt{3}{\bm \rho} \cdot {\bm \lambda})}\right)
\label{hypVY1d} \\
&& \textrm{when}  \left\{
\begin{array}{l}
\quad 2 {\bm \rho}^2 - \sqrt{3}{\bm \rho} \cdot {\bm \lambda} \leq
- \rho \sqrt{{\bm \rho}^2 + 3 {\bm \lambda}^2
- 2 \sqrt{3}{\bm \rho} \cdot {\bm \lambda}} \\
\quad 2 {\bm \rho}^2 + \sqrt{3}{\bm \rho} \cdot {\bm \lambda} \geq
- \rho \sqrt{{\bm \rho}^2 + 3 {\bm \lambda}^2 + 2 \sqrt{3}{\bm
\rho} \cdot {\bm \lambda}}  \nonumber \\
\quad 3 {\bm \lambda}^2 - {\bm \rho}^2 \geq - \frac{1}{2} \sqrt{
({\bm \rho}^2 + 3 {\bm \lambda}^2)^{2} - 12 ({\bm \rho} \cdot {\bm
\lambda})^{2}}
\end{array} \right. \, .
\end{eqnarray}
\end{subequations}
\twocolumngrid
The O(6) $v_{6 \pm 6}$ coefficient is defined in Eq. (\ref{e:def_ME})
\begin{equation}
v_{6,\pm 6}^{\rm Y-string} =
\int~ {\cal Y}^{6\pm6}_{00}(\Omega_{5}) \,  V_{\rm Y-string}
(\alpha, \phi)\, \, d\Omega_{(5)}
\label{e:def_66}
\end{equation}
where the integration over $d\Omega_{(5)}$ is constrained by inequalities
(\ref{hypVY1b}) - (\ref{hypVY1d}) and  
\begin{eqnarray}
{\mathcal Y}^{6,\pm6}_{00}(\alpha, \phi) &=& \frac{2}{\pi^{3/2}} R^{-6}
\left({\bm \lambda}^2 - {\bm \rho}^2 \pm 2 i {\bm \lambda} \cdot {\bm \rho} \right)^3
\nonumber \\
&=&  \frac{\mp 2 i }{\pi^{3/2}}  \sin^3 \alpha \exp \left(\mp 3 i \phi \right) \
\end{eqnarray}
which is equivalent, up to the normalization constant, to the O(3) spherical harmonics
$Y_{3,\pm 3}\left(\alpha,\phi \right)$.
Numerical evaluation yields $v_{6, \pm 6}^{\rm Y-string} = -0.027$, which value is
smaller than the subsequent coefficients in the expansion of this potential, see
Table \ref{tab:Potential_expansion}.

\section{Details of calculations}
\label{a:calculations}

\subsection{$\K=2$ shell}
\label{a:K12}

The calculated coefficients entering the effective potentials for states with $\K = 2$
can be found in Table \ref{tab:drcc2n2}.
\begin{table}[tbh]
\begin{center}
\caption{The values of the three-body potential hyper-angular matrix elements
$\pi \sqrt{\pi} \langle \, {\mathcal Y}^{4,0}_{00} \, \rangle_{\rm ang}$,
for various K = 2 states (for all allowed orbital waves L). The correspondence
between the $S_3$ permutation group irreps. and SU(6)$_{FS}$ symmetry multiplets
of the three-quark system: $S \leftrightarrow 56$, $A \leftrightarrow 20$
and $M \leftrightarrow 70$.}
\begin{tabular}{cccc}
\hline \hline K & $\left(K, Q, L, M, \nu \right)$ &  $[SU(6),L^P]$ & $\pi \sqrt{\pi} \langle \,{\mathcal Y}^{4,0}_{00}\,
\rangle_{\rm ang}$ \\
\hline
\hline
2 & $\left(2, -2, 0, 0, 0\right)$ & $[70,0^+]$ & $\frac{1}{\sqrt{3}}$ \\ 
2 & $\left(2, ~0, 2, 2, 0\right)$ & $[56,2^+]$ & $\frac{\sqrt{3}}{5}$ \\ 
2 & $\left(2, \mp 2, 2, 2, \pm 3\right)$ & $[70,2^+]$ & $-\frac{1}{5\sqrt{3}}$ \\ 
2 & $\left(2, 0, 1, 1, 0\right)$ & $[20,1^+]$ & $-\frac{1}{\sqrt{3}}$ \\ 
\hline
\end{tabular}
\label{tab:drcc2n2}
\end{center}
\end{table}

\subsection{$\K=3$ shell}
\label{a:K3}

The calculated effective potentials in states with of K=3 and various values L are listed in
Tables \ref{tab:drcc3nSU3} and \ref{tab:drcc34b}.
\begin{table}[tbh]
\begin{center}
\caption{The values of
the three-body potential hyper-angular diagonal matrix elements
$\langle \, {\mathcal Y}^{4,0}_{00} \,\rangle_{\rm ang}$,
for various K = 3 states (for all allowed orbital waves L).}
\begin{tabular}{cccc}
\hline \hline K
& $\left(K, Q, L, M, \nu \right)$ & $[SU(6),L^P]$ & $\pi \sqrt{\pi} \langle \, {\cal Y}^{4,0}_{00} \,\rangle_{\rm ang}$ \\
\hline
\hline
3 & $\left(3, \mp 3, ~1, ~1, \pm 1\right)$ & $[20,1^-]$ & $~\frac{1}{\sqrt{3}}$ \\ 
3 & $\left(3, \mp 3, ~1, ~1, \pm 1\right)$ & $[56,1^-]$ & $~\frac{1}{\sqrt{3}}$ \\ 
3 & $\left(3, \pm 1, ~1, ~1, \pm 3\right)$ & $[70,1^-]$ & $~0$ \\ 
3 & $\left(3, \mp 1, ~2, ~2, \pm 5\right)$ &  $[70,2^-]$ & $-\frac{1}{\sqrt{3}}$ \\ 
3 & $\left(3, \mp 1, ~3, ~3, \pm 2\right)$ & $[70,3^-]$ & $~\frac{5}{7 \sqrt{3}}$ \\ 
3 & $\left(3, \pm 3, ~3, ~3, \mp 6\right)$ & $[56,3^-]$ & $-\frac{\sqrt{3}}{7}$ \\ 
3 & $\left(3, \pm 3, ~3, ~3, \mp 6\right)$ & $[20,3^-]$ & $-\frac{\sqrt{3}}{7}$ \\ 
\hline
\end{tabular}
\label{tab:drcc3nSU3}
\end{center}
\end{table}
\begin{table}[tbh]
\begin{center}
\caption{The values of the off-diagonal matrix elements
of the hyper-angular part of the three-body potential
$\pi \sqrt{\pi} \langle [SU(6)_{f},L_{f}^P] |\,2 \Re e \mathcal{Y}^{6, \pm 6, 0}_{0, 0}
\,| [SU(6)_{i},L_{i}^P] \rangle_{\rm ang}$,
for various K = 3 states (for all allowed orbital waves L).}
\begin{tabular}{cccc}
\hline \hline K & $[SU(6)_{f},L_{f}^P]$
& $[SU(6)_{i},L_{i}^P]$ & $\pi \sqrt{\pi} \langle 2 \Re e
\mathcal{Y}^{6, \pm 6, 0}_{0, 0} \rangle_{\rm ang}$ \\
\hline
\hline
3 & $[20,1^-]$ & $[20,1^-]$ & $-1$ \\ 
3 & $[56,1^-]$ & $[56,1^-]$ & $~~1$ \\ 
3 & $[20,3^-]$ & $[20,3^-]$ & $-\frac{2}{7}$ \\ 
3 & $[56,3^-]$ & $[56,3^-]$ & $~~\frac{2}{7}$ \\ 
3 & $[70,L^-]$ & $[70,L^-]$ & $~~0$ \\ 
\hline
\end{tabular}
\label{tab:drcc34b}
\end{center}
\end{table}

\subsection{$\K=4$ shell}
\label{a:K4}

The calculated effective potentials for states with K=4 and various values of L are listed in
Table 
\ref{tab:drcc4n2d}.
\begin{table}[tbh]
\begin{center}
\caption{The values of
the three-body potential hyper-angular diagonal matrix elements
$\langle \, {\mathcal Y}^{4,0,0}_{00} \,\rangle_{\rm ang}$
and
$\langle \, {\mathcal Y}^{8,0,0}_{00} \,\rangle_{\rm ang}$,
for various K = 4 states (for all allowed orbital waves L).}
\begin{tabular}{ccccc}
\hline \hline K
& $\left(K, Q, L, M, \nu \right)$ & $[SU(6),L^P]$ & $\pi \sqrt{\pi} \langle \, {\mathcal Y}^{4,0,0}_{00} \,\rangle_{\rm ang}$
& $\pi \sqrt{\pi} \langle \, {\mathcal Y}^{8,0,0}_{00} \,\rangle_{\rm ang}$ \\
\hline
\hline
4 & $\left(4, \pm 4, 0, 0, ~~0 \right)$ & $[70,0^+]$ & $~\frac{\sqrt{3}}{2}$ & $~\frac{1}{2\sqrt{5}}$ \\
4 & $\left(4, ~~0, 0, 0, ~~0 \right)$ & $[56,0^+]$ & $~0$ & $~\frac{2}{\sqrt{5}}$ \\
4 & $\left(4, \pm 2, 1, 1, \pm 2 \right)$ & $[70,1^+]$ & $~~0$ & $-\frac{1}{\sqrt{5}}$ \\
4 & $\left(4, 0, 2, 2, \mp \sqrt{105}\right)$ & $[56,2^+]$ & $-\frac{12 \sqrt{3}}{35}$ & $\frac{\sqrt{5}}{7}$ \\
4 & $\left(4, 0, 2, 2, \mp \sqrt{105}\right)$ & $[20,2^+]$ & $0$ & $-\frac{1}{\sqrt{5}}$ \\
4 & $\left(4, \pm 2, 2, 2, \pm 2 \right)$ & $[70,2^+]$ & $\frac{4 \sqrt{3}}{35}$ & $\frac{\sqrt{5}}{7}$ \\
4 & $\left(4, \pm 4, 2, 2, \mp 3 \right)$ & $[70^{'},2^+]$ & $~\frac{2 \sqrt{3}}{7}$ & $-\frac{1}{7 \sqrt{5}}$ \\
4 & $\left(4, \mp 2, 3, 3, \pm 13 \right)$ & $[70,3^+]$ & $-\frac{5 \sqrt{3}}{14}$ & $~\frac{1}{14 \sqrt{5}}$ \\
4 & $\left(4, ~~0, 3, 3, ~~0 \right)$ & $[20,3^+]$ & $-\frac{3 \sqrt{3}}{14}$ & $-\frac{\sqrt{5}}{14}$ \\
4 & $\left(4, ~~0, 4, 4, ~~0 \right)$ & $[56,4^+]$ & $\frac{5 \sqrt{3}}{14}$ & $~\frac{3}{14 \sqrt{5}}$ \\
4 & $\left(4, \mp 2, 4, 4, \pm 5 \right)$ & $[70,4^+]$ & $\frac{3 \sqrt{3}}{14}$ & $-\frac{\sqrt{5}}{42}$ \\
4 & $\left(4, \mp 4, 4, 4, \pm 10 \right)$ & $[70^{'},4^+]$ & $-\frac{3 \sqrt{3}}{14}$  & $\frac{1}{42 \sqrt{5}}$ \\
\hline
\end{tabular}
\label{tab:drcc4n2d}
\end{center}
\end{table}
\begin{table}[tbh]
\begin{center}
\caption{The values of the off-diagonal matrix elements of the hyper-angular part of the three-body potential
$\pi \sqrt{\pi} \langle [SU(6)_{f},L_{f}^P] |\, 2 \Re e \mathcal{Y}^{6, \pm 6, 0}_{0, 0}
\,| [SU(6)_{i},L_{i}^P] \rangle_{\rm ang}$,
for various K = 4 states (for all allowed orbital waves L).}
\begin{tabular}{cccc}
\hline \hline K & $[SU(6)_{f},L_{f}^P]$
& $[SU(6)_{i},L_{i}^P]$ & $\pi \sqrt{\pi} \langle  2 \Re e \mathcal{Y}^{6, \pm 6, 0}_{0, 0}  \rangle_{\rm ang}$ \\
\hline
\hline
4 & $[70,2^+]$ & $[70^{'},2^+]$ & $\frac{6}{7} \sqrt{\frac{6}{5}}$ \\ 
4 & $[70^{'},2^+]$ & $[70,2^+]$ & $\frac{6}{7} \sqrt{\frac{6}{5}}$ \\ 
4 & $[70,4^+]$ & $[70^{'},4^+]$ & $\frac{8}{21}$ \\ 
4 & $[70^{'},4^+]$ & $[70,4^+]$ & $\frac{8}{21}$ \\ 
\hline
4 & $[20,L^+]$ & $[20,L^+]$ & $~~0$ \\ 
4 & $[56,L^+]$ & $[56,L^+]$ & $~~0$ \\ 
4 & $[20,L^+]$ & $[56,L^+]$ & $~~0$ \\ 
\hline
\end{tabular}
\label{tab:drcc44b}
\end{center}
\end{table}
The selection rules that we have not derived fully, as yet, are:
1) the 3D expansion of the potentials goes in double-valued steps of $K$ and $\G$,
as compared with the 2D case: {\it viz.} : $K=0,4,8,12$ and $K=6,\G=6$ in 3D, and $K=0,2,4,6$
and $K=3,\G=3$ in 2D. The latter can be understood
in terms of O(3) Clebsch-Gordan coefficients and spherical harmonics, whereas the former
can be understood in terms of O(6) Clebsch-Gordan coefficients, whose properties are not
(well) known, however.
2) the selection rules read: $\G \equiv$ 0 (mod 6) and $K \equiv$ 0 (mod 4);
and that the Clebsch-Gordan coefficients demand 
$\G = |\G_f - \G_i|$.

The $\phi$-dependent (``two-body") component in the three-body potential, which is proportional to
$v_{6\pm 6}$, enters the $K = 4$ spectrum, 
only through the off-diagonal matrix elements of two pairs of mixed-symmetry $[70,L^P]$-plets:
The multiplet states $| [70,L^+] \rangle$ and $| [70^{'},L^+] \rangle$ have identical physical
quantum numbers $(K,L^P)$, whereas the democracy label $\G$ is generally not a good quantum
number in permutation-symmetric three-body potentials, so it may be expected to be broken,
and the corresponding eigenstates to mix under the influence of general permutation-symmetric
three-body potentials. That is precisely what happens when the expansion coefficients $v_{6 \pm 6} \neq 0$
does not vanish. 
In that case the two multiplets $| [70, L^+] \rangle$ and $| [70^{'}, L^+] \rangle$ mix, as determined
by the diagonalization of the 2$\times$2 potential matrix.

\subsubsection{$| [70, L^+] \rangle - | [70^{'}, L^+] \rangle$ mixing and the physical states}
\label{ss:mixing}

The three-body potential matrix in the O(6) symmetric states basis is non-diagonal in general;
for example, for two multiplets ($|{\rm a} \rangle$, $|{\rm b} \rangle$) that have identical quantum
numbers, such as $| [70, L^+] \rangle$ and $| [70^{'}, L^+] \rangle$, the potential matrix is $2 \times 2$
and can be written as
\onecolumngrid
\begin{eqnarray}
V_{\rm a,b} = \frac{1}{\pi \sqrt{\pi}}
\left(\begin{array}{cc}
v_{0 0} + \left[v_{4 0} \langle \, {\mathcal Y}^{4,0,0}_{00} \,\rangle_{\rm a}
+ v_{8 0} \langle \, {\mathcal Y}^{8,0,0}_{00} \,\rangle_{\rm a} \right]
& v_{6\pm 6} \langle  2 \Re e \mathcal{Y}^{6, \pm 6, 0}_{0, 0}  \rangle_{\rm a,b} \\
v_{6\pm 6} \langle  2 \Re e \mathcal{Y}^{6, \pm 6, 0}_{0, 0}  \rangle_{\rm b,a} &
v_{0 0} + \left[v_{4 0} \langle \, {\mathcal Y}^{4,0,0}_{00} \,\rangle_{\rm b}
+ v_{8 0} \langle \, {\mathcal Y}^{8,0,0}_{00} \,\rangle_{\rm b}
\right]
\end{array}\right) 
~\
\label{e:massm}
\end{eqnarray}
\twocolumngrid
where $v_{0 0}, v_{4 0}, v_{8 0} $ and $v_{6\pm 6}$ are the hyperspherical expansion coefficients
of the potential in question and $\langle \, {\mathcal Y}^{K,0,0}_{00} \,\rangle_{\rm a}$,
$\langle \, {\mathcal Y}^{K,0,0}_{00} \,\rangle_{\rm b}$ are the $K$-th diagonal hyper-angular matrix
elements for SO(6) state $|{\rm a} \rangle$, $|{\rm b} \rangle$, respectively,
that can be read off from Table \ref{tab:drcc4n2d}, and
$\langle  2 \Re e \mathcal{Y}^{6, \pm 6, 0}_{0, 0}  \rangle_{\rm a,b}$ is the off-diagonal matrix
element, from Table \ref{tab:drcc44b}.
Diagonalization is accomplished by way of mixing of the
$|{[70, L^+]}_{\rm a} \rangle$, and $|{[70, L^+]}_{\rm b} \rangle$ states,
\begin{eqnarray}
| [70~, L^+] \rangle &=&
~~\cos \theta |{[70~, L^+]}_{\rm a} \rangle +
\sin \theta |{[70^{'}, L^+]}_{\rm b} \rangle,
\nonumber \\
| [70^{'} , L^+] \rangle &=&
- \sin \theta |{[70~, L^+]}_{\rm a} \rangle +
\cos \theta |{[70^{'}, L^+]}_{\rm b} \rangle, \
\label{e:thetaS} \
\end{eqnarray}
the mixing angle $\theta$ being determined by
\onecolumngrid
\begin{eqnarray}
\tan 2 \theta &=&
\frac{2 v_{6\pm 6} \langle  2 \Re e \mathcal{Y}^{6, \pm 6, 0}_{0, 0}
\rangle_{\rm a,b} }{\left[v_{4 0} \langle \, {\mathcal Y}^{4,0,0}_{00} \,\rangle_{\rm a}
+ v_{8 0} \langle \, {\mathcal Y}^{8,0,0}_{00} \,\rangle_{\rm a} \right] -
\left[v_{4 0} \langle \, {\mathcal Y}^{4,0,0}_{00} \,\rangle_{\rm b}
+ v_{8 0} \langle \, {\mathcal Y}^{8,0,0}_{00} \,\rangle_{\rm b} \right]} ~. \
\label{eigenmassc}
\end{eqnarray}
The (diagonal) eigen-values of the potential matrix
\begin{eqnarray}
V_{\rm a,b} =  
\left(\begin{array}{cc}
a & c \\
c & d \end{array}\right) 
~\
\label{e:matrix}
\end{eqnarray}
can also be expressed in terms of the matrix elements $(a,c,d)$ as follows,
\[V_{\pm} = \frac{1}{2} \left(a+d \pm \sqrt{a^2-2 a d+4 c^2+d^2}\right),\]
and that leads to, for the $[70,4]$-plets
\[ 
V_{\pm}([70,4]) =  \frac{1}{\pi \sqrt{\pi}} \left(v_{0 0} + \frac{1}{42 \sqrt{5}} \left(- 2 v_{8 0} \pm
\sqrt{1215 v_{4 0}^2 - 54 \sqrt{15} v_{4 0} v_{8 0} + 9 v_{8 0}^2 + 1280 v_{6\pm 6}^2} \right)\right)\]
and for the $[70,2]$-plets
$$
V_{\pm}([70,2]) =  \frac{1}{\pi \sqrt{\pi}} \left(v_{0 0} +
\frac{1}{35} \left(7 \sqrt{3} v_{4 0} + 2 \sqrt{5} v_{8 0} \pm
3 \sqrt{3 v_{4 0}^2 - 2 \sqrt{15} v_{4 0} v_{8 0} + 5 v_{8 0}^2 + 120 v_{6\pm 6}^2}\right)\right)$$
where $b = v_{4 0}$, $c = v_{8 0}$ and $d = v_{6\pm 6}$.
\twocolumngrid

\subsection{$\K=5$ shell}
\label{a:K5}

The calculated effective potentials of states with K=5 and various values of L are listed in
Tables \ref{tab:result5}, \ref{tab:drcc5c}, \ref{tab:drcc54b}. 
\onecolumngrid
\begin{table}[tbh]
\begin{center}
\caption{The values of the three-body potential hyper-angular diagonal matrix elements
$\langle \, {\mathcal Y}^{4,0,0}_{00} \,\rangle_{\rm ang}$,
$\langle \, {\mathcal Y}^{8,0,0}_{00} \,\rangle_{\rm ang}$, and
$\pi \sqrt{\pi} \langle \, 2 \Re e{\mathcal Y}^{6,\pm 6,0}_{00} \,\rangle_{\rm ang}$
for various K = 5 SU(6) multiplets (with orbital angular momentum $L=J$). States containing one,
or more asterices ($*$) are subject to mixing described in the text.}
\begin{tabular}{cccccc}
\hline \hline K
& $\left(K, Q, L, M, \nu \right)$ & $[SU(6),L^P]$ & $\pi \sqrt{\pi} \langle \,
{\mathcal Y}^{4,0,0}_{00} \,\rangle_{\rm ang}$
& $\pi \sqrt{\pi} \langle \, {\mathcal Y}^{8,0,0}_{00} \,\rangle_{\rm ang}$ &
$\pi \sqrt{\pi} \langle \, 2 \Re e{\mathcal Y}^{6,\pm 6,0}_{00} \,\rangle_{\rm ang}$ \\
\hline
\hline
5 & $\left(5, -5, 1, 1, 1 \right)$ & $[70,1^-]$ & $~\frac{\sqrt{3}}{2}$ & $~\frac{1}{2 \sqrt{5}}$ & $~~*$ \\
5 & $\left(5, -1, 1, 1, 3 \right)$ & $[70^{'},1^-]$ & $~0$ & $~\frac{1}{\sqrt{5}}$ & $~~*$ \\
5 & $\left(5, -3, 1, 1, -5 \right)$ & $[56,1^-]$ & $~\frac{1}{2 \sqrt{3}}$ & $-\frac{1}{2 \sqrt{5}}$ & $-\frac{2}{5}$ \\
5 & $\left(5, -3, 1, 1, -5 \right)$ & $[20,1^-]$ & $~\frac{1}{2 \sqrt{3}}$ & $-\frac{1}{2 \sqrt{5}}$ & $~~\frac{2}{5}$ \\
5 & $\left(5, -1, 2, 2, -13 \right)$ & $[70,2^-]$ & $-\frac{1}{2 \sqrt{3}}$ & $-\frac{1}{2 \sqrt{5}}$ & $~~0$ \\
5 & $\left(5, -3, 2, 2, 3 \right)$ & $[56,2^-]$ & $~0$ & $-\frac{1}{\sqrt{5}}$ & $-\frac{3}{5}$ \\
5 & $\left(5, -3, 2, 2, 3 \right)$ & $[20,2^-]$ & $0$ & $-\frac{1}{\sqrt{5}}$ & $~~\frac{3}{5}$ \\
5 & $\left(5, -5, 3, 3, 6 \right)$ & $[70,3^-]$ & $~\frac{2}{3 \sqrt{3}}$ & $-\frac{1}{3 \sqrt{5}}$ & $~**$ \\
5 & $\left(5, -1, 3, 3, 7 - \sqrt{241} \right)$ & $[70^{'},3^-]$ &
$-\frac{5}{12 \sqrt{3}} + \frac{85}{12 \sqrt{723}}$ &
$\frac{241 + 19 \sqrt{241}}{2892 \sqrt{5}}$
& $~**$ \\
5 & $\left(5, -1, 3, 3, 7 + \sqrt{241}\right)$ & $[70^{''},3^-]$ &
$- \frac{5 (241 + 17 \sqrt{241})}{2892 \sqrt{3}}$ &
$\frac{241 - 19 \sqrt{241}}{2892 \sqrt{5}}$
& $~**$ \\
5 & $\left(5, -3, 3, 3, 0 \right)$ & $[56,3^-]$ & $~\frac{1}{2 \sqrt{3}}$ & $~\frac{7}{6 \sqrt{5}}$ & $~~\frac{14}{15}$ \\
5 & $\left(5, -3, 3, 3, 0 \right)$ & $[20,3^-]$ & $~\frac{1}{2 \sqrt{3}}$ & $~\frac{7}{6 \sqrt{5}}$ & $-\frac{14}{15}$ \\
5 & $\left(5, -1, 4, 4, 8 \right) $ & $[70,4^-]$ & $-\frac{1}{2 \sqrt{3}}$ & $-\frac{1}{2 \sqrt{5}}$ & $~~0$ \\
5 & $\left(5, -3, 4, 4, 24 \right) $ & $[56,4^-]$ & $-\frac{7}{6 \sqrt{3}}$ & $~\frac{1}{6 \sqrt{5}}$ & $-\frac{2}{15}$ \\
5 & $\left(5, -3, 4, 4, 24 \right)$ & $[20,4^-]$ & $-\frac{7}{6 \sqrt{3}}$ & $~\frac{1}{6 \sqrt{5}}$ & $~~\frac{2}{15}$ \\
5 & $\left(5, -5, 5, 5, 15 \right)$ & $[70,5^-]$ & $-\frac{5}{6 \sqrt{3}}$ & $\frac{\sqrt{5}}{66}$ & $***$ \\
5 & $\left(5, -1, 5, 5, 3 \right)$ & $[70^{'},5^-]$ & $~\frac{7}{6 \sqrt{3}}$ & $~\frac{17}{66 \sqrt{5}}$ & $***$ \\
5 & $\left(5, -3, 5, 5, 9 \right)$ & $[56,5^-]$ & $~\frac{1}{2 \sqrt{3}}$ & $-\frac{19}{66 \sqrt{5}}$ & $~~\frac{8}{15}$ \\
5 & $\left(5, -3, 5, 5, 9 \right)$ & $[20,5^-]$ & $~\frac{1}{2 \sqrt{3}}$ & $-\frac{19}{66 \sqrt{5}}$ & $-\frac{8}{15}$ \\
\hline
\end{tabular}
\label{tab:result5}
\end{center}
\end{table}

The $\phi$-dependent (``two-body") potential component proportional to
$v_{6\pm 6}$ enters these effective potentials in two ways:
1) through diagonal matrix elements in Table \ref{tab:drcc5c},
causing the splitting of symmetric $[56,L^P]$ and antisymmetric $[20,L^P]$ multiplets,
as in the $K=3$ case;
2) through off-diagonal matrix elements in Table \ref{tab:drcc54b}, causing further splitting
of two mixed-symmetry $[70,L^P]$-plets, as in the $K=4$ case.
\begin{table}[tbh]
\begin{center}
\caption{The values of the diagonal matrix elements
of the hyper-angular part of the three-body potential
$\langle \mathcal{Y}\left(K, Q_f, L, M, \nu_{f} \right) |\,
2 \Re e \mathcal{Y}^{6, \pm 6, 0}_{0, 0} \,|
\mathcal{Y}\left(K, Q_i, L, M, \nu_{i} \right) \rangle_{\rm ang}$,
for various K = 5 states (for all allowed orbital waves L).}
\begin{tabular}{cccc}
\hline \hline K & $[SU(6)_{f},L_{f}^P]$ & $[SU(6)_{i},L_{i}^P]$
& $\pi \sqrt{\pi} \langle 2 \Re e \mathcal{Y}^{6, \pm 6, 0}_{0, 0} \rangle_{\rm ang}$ \\
\hline
\hline
5 & $
[56,1^-]$ & $
[56,1^-]$ & $-\frac{2}{5}$ \\
5 & $
[20,1^-]$ & $
[20,1^-]$ & $~~\frac{2}{5}$ \\
5 & 
$[56,2^-]$ & 
$[56,2^-]$ & $~~\frac{3}{5}$ \\
5 & 
$[20,2^-]$ & 
$[20,2^-]$ & $-\frac{3}{5}$ \\
5 & 
$[56,3^-]$ & 
$[56,3^-]$ & $~~\frac{14}{15}$ \\
5 & $[20,3^-]$ &$[20,3^-]$ & $-\frac{14}{15}$ \\
5 & 
$[56,4^-]$ & 
$[56,4^-]$ & $-\frac{2}{15}$ \\
5 & 
$[20,4^-]$ & 
$[20,4^-]$ & $~~\frac{2}{15}$ \\
5 & $
[56,5^-]$ & $
[56,5^-]$ & $~~\frac{8}{15}$ \\
5 & $
[20,5^-]$ & $
[20,5^-]$ & $-\frac{8}{15}$ \\
\hline
\end{tabular}
\label{tab:drcc5c}
\end{center}
\end{table}
\begin{table}[tbh]
\begin{center}
\caption{The values of the off-diagonal matrix elements of the hyper-angular part of the three-body potential
$\pi \sqrt{\pi} \langle [SU(6)_{f},L_{f}^P] |\, 2 \Re e \mathcal{Y}^{6, \pm 6, 0}_{0, 0}
\,| [SU(6)_{i},L_{i}^P] \rangle_{\rm ang}$,
for various K = 5 states (for all allowed orbital waves L).}
\begin{tabular}{cccc}
\hline \hline K & $[SU(6)_{f},L_{f}^P]$
& $[SU(6)_{i},L_{i}^P]$ & $\pi \sqrt{\pi} \langle
2 \Re e \mathcal{Y}^{6, \pm 6, 0}_{0, 0} \rangle_{\rm ang}$ \\
\hline
\hline
5 & $[70,1^-]$ & $[70^{'},1^-]$ & $\frac{\sqrt{6}}{5}$ \\ 
5 & $[70^{'},1^-]$ & $[70,1^-]$ & $\frac{\sqrt{6}}{5}$ \\ 
5 & $[70,3^-]$ & $[70^{'},3^-]$ & $\sqrt{\frac{139}{450}+\frac{2131}{450 \sqrt{241}}}$ \\ 
5 & $[70^{'},3^-]$ & $[70,3^-]$ & $\sqrt{\frac{139}{450}+\frac{2131}{450 \sqrt{241}}}$ \\ 
5 & $[70,3^-]$ & $[70^{''},3^-]$ & $-\frac{1}{15} \sqrt{\frac{1}{482} \left(33499-2131 \sqrt{241}\right)}$ \\ 
5 & $[70^{''},3^-]$ & $[70,3^-]$ & $-\frac{1}{15} \sqrt{\frac{1}{482} \left(33499-2131 \sqrt{241}\right)}$ \\ 
5 & $[70,5^-]$ & $[70^{'},5^-]$ & $\frac{2}{3} \sqrt{\frac{2}{5}}$ \\ 
5 & $[70^{'},5^-]$ & $[70,5^-]$ & $\frac{2}{3} \sqrt{\frac{2}{5}}$ \\ 
\hline
\end{tabular}
\label{tab:drcc54b}
\end{center}
\end{table}
\twocolumngrid
Just as in Sect. \ref{ss:mixing}, the three-body potential matrix in the O(6) symmetric states
basis is non-diagonal in general, and can be diagonalized in the same manner.

\subsubsection{Two-state $| [70, L^P] \rangle - | [70^{'}, L^P] \rangle$ mixing}

Diagonalization of the $2 \times 2$ matrices proceeds 
by way of mixing of the $|{[70, L^+]}_{\rm a} \rangle$, and $|{[70, L^+]}_{\rm b} \rangle$ states,
as determined by Eq. (\ref{e:thetaS}), and
the mixing angle $\theta$ being given by Eq. (\ref{eigenmassc}).
The (diagonal) eigen-values of the potential matrix Eq. (\ref{e:matrix})
can also be expressed in terms of the matrix elements 
and that leads to, for the $[70,5^-]$-plets, see Table \ref{tab:drcc54b},
\onecolumngrid
\[V_{\pm}([70,5]) =  \frac{1}{\pi \sqrt{\pi}} \Bigg(v_{0 0} + \frac{\sqrt{3}}{18} v_{4 0}
+ \frac{\sqrt{5}}{30} v_{8 0} + \]
\[\pm \frac{\sqrt{5}}{165} \sqrt{1815 v_{4 0}^2
+ 66 \sqrt{15} v_{4 0} v_{8 0} + 9 v_{8 0}^2 + 968 v_{6\pm 6}^2} \Bigg)\]
and for the $[70,1^-]$-plets, see Table \ref{tab:drcc54b},
\[V_{\pm}([70,1]) =  \frac{1}{\pi \sqrt{\pi}} \Bigg( v_{0 0} + \frac{\sqrt{3}}{4} v_{4 0}  +
\frac{3 \sqrt{5}}{20} v_{8 0}\]
\[\pm \frac{1}{20} \sqrt{75 v_{4 0}^2 - 10 \sqrt{15} v_{4 0} v_{8 0} + 5 v_{8 0}^2 + 96 v_{6\pm 6}^2} \Bigg)\]
where $b = v_{4 0}$,
$c = v_{8 0}$ and
$d = v_{6\pm 6}$.

\subsubsection{Three-state $| [70, 3^-] \rangle - | [70^{'}, 3^-] \rangle  - | [70^{''}, 3^-] \rangle$ mixing}

In the $L =3$ case the mixing potential matrix is $3 \times 3$, see Table \ref{tab:drcc54b}:
\begin{eqnarray}
V_{\rm a,b} =  
\left(\begin{array}{ccc}
\alpha & \delta & 0 \\
\delta & \beta & \epsilon \\
0 & \epsilon & \gamma \end{array}\right) 
~\
\label{e:matrix3}
\end{eqnarray}
Its eigenvalues 
can also be expressed in terms of the matrix elements $(\alpha, \beta, \gamma, \delta, \epsilon)$ as follows,
and that leads to the following potential eigenvalues, for the $[70,3^-]$-plets,
\begin{eqnarray}
V([ 70, 3^-]) &=
\frac{1}{3} (\alpha + \beta + \gamma) +~~~ \frac{1}{3 \sqrt[3]{2}} A 
&- \frac{\sqrt[3]{2}}{3 A} I 
 \label{e:K5_70a_3} \\
V([70^{'},3^-]) &=
\frac{1}{3} (\alpha + \beta + \gamma)
- \frac{\left(1 - i \sqrt{3}\right)}{6 \sqrt[3]{2}} A
&+ \frac{\left(1 + i \sqrt{3}\right)}{3~ \sqrt[3]{4} A} I
\label{e:K5_70b_3} \\
V([70^{''},3^-]) &=
\frac{1}{3}(\alpha + \beta + \gamma)
- \frac{\left(1 - i \sqrt{3}\right)}{6 \sqrt[3]{2}}A 
&+ \frac{\left(1 + i \sqrt{3}\right)}{3~ \sqrt[3]{4} A}I
 \label{e:K5_70c_3} \
\end{eqnarray}
where
$C$ and $D$ have been separated into the unperturbed ($\epsilon = \delta = 0$) part and the
perturbation - collect the $\delta^2 + \epsilon^2$ terms together:
\begin{eqnarray}
C &=& 2 \alpha^3 + 2 \beta^3 + 2 \gamma^3 - 3 \left(\beta  + \gamma \right) \left(\beta \gamma
+ \alpha^2 \right) - 3 \left(\beta^2 + \gamma^2 \right) \alpha
+ 12 \alpha \beta \gamma \nonumber \\
&+& 9 \left[\left( \delta^2 - 2 \epsilon^2 \right) \alpha + \gamma \left(\epsilon^2
- 2 \delta^2 \right) + \beta \left(\epsilon^2 + \delta^2 \right) \right]
\label{e:C} \\
D &=&
4 I^3  + C^2
\label{e:D}  \\
I &=& \left(-\alpha^2- \beta^2 -\gamma^2  + \beta \alpha + \gamma \alpha + \beta  \gamma \right)
- 3 \left(\delta^2 + \epsilon^2 \right)
\label{e:I}\
\end{eqnarray}
Here
\begin{eqnarray}
\alpha =& \frac{1}{\pi \sqrt{\pi}}
\Bigg( & v_{00}
- \left( \frac{5}{12 \sqrt{3}} - \frac{85}{12 \sqrt{723}} \right) v_{40}
+ \frac{241 + 19 \sqrt{241}}{2892 \sqrt{5}} v_{80} \Bigg) \nonumber \\
\beta =& \frac{1}{\pi \sqrt{\pi}} \Bigg(& v_{00} +
~~~~~~~~~~~~~~~~~~~~\frac{2}{3\sqrt{3}} v_{40} + ~~~~~~~~~
- \frac{1}{3 \sqrt{5}} v_{80} \Bigg) \nonumber \\
\gamma = & \frac{1}{\pi \sqrt{\pi}} \Bigg(& v_{00}
- \frac{5 (241 - 17 \sqrt{241})}{2892 \sqrt{3}} v_{40} ~~~~~
+ \frac{241 - 19 \sqrt{241}}{2892 \sqrt{5}} v_{80} \Bigg) \nonumber \\
\delta =& \frac{1}{\pi \sqrt{\pi}}
\Bigg( & \sqrt{\frac{139}{450}+\frac{2131}{450 \sqrt{241}}}
v_{66}  \Bigg) \nonumber \\
\epsilon = & \frac{1}{\pi \sqrt{\pi}} \Bigg(&
-\frac{1}{15} \sqrt{\frac{1}{482} \left(33499-2131 \sqrt{241}\right)}
v_{66} \Bigg) \nonumber \
\end{eqnarray}
These formulae are manifestly rather cumbersome, and they do not offer much 
new insight into the problem that could
not be gained by a (simpler) numerical calculation.
Clearly, there is no advantage to having explicit algebraic expressions for this kind of quantity.
As $K$ increases to $K \geq 6$, the number of mixing multiplets can only increase, as can the number of
states within invariant sub-spaces.
\twocolumngrid

\end{document}